\documentclass[a4paper,11pt]{article}
\usepackage{jcappub} %

\usepackage{graphicx}	%
\usepackage{latexsym,amssymb}
\usepackage{amsmath}
\usepackage{float}
\usepackage{amsmath,morefloats}
\usepackage{natbib}
\usepackage{xcolor}
\usepackage{color}
\usepackage{verbatim}
\usepackage{xspace}
\usepackage{tikz}
\usepackage{mathabx}

\usepackage{my_commands}
\usepackage{soul}

\graphicspath{{figures/}}

\arxivnumber{1234.56789} %
\title{Relationship between 2D and 3D Galaxy Stellar Mass and Correlations with Halo Mass}

\author[a,b]{Conghao Zhou,}
\author[c]{Alexie Leauthaud,}
\author[d]{Shuo Xu,}
\author[e]{Benedikt Diemer,}
\author[d]{Song Huang,}
\author[e]{Katya Leidig,}
\author[a,b]{Tesla Jeltema,}
\author[f]{Marco Gatti,}
\author[c,g]{Yifei Luo,}
\author[h,i]{Carlo Cannarozzo,}
\author[c]{Sven Heydenreich.}

\affiliation[a]{Physics Department, University of California, Santa Cruz, CA 95064, USA}
\affiliation[b]{Santa Cruz Institute for Particle Physics, Santa Cruz, CA 95064, USA}
\affiliation[c]{Department of Astronomy and Astrophysics, UCO/Lick Observatory, University of California, 1156 High Street, Santa Cruz, CA 95064, USA}
\affiliation[d]{Department of Astronomy, Tsinghua University, Beijing 100084, China}
\affiliation[e]{Department of Astronomy, University of Maryland, College Park, MD 20742, USA}
\affiliation[f]{Kavli Institute for Cosmological Physics, University of Chicago, Chicago, IL 60637, USA}
\affiliation[g]{Lawrence Berkeley National Laboratory, One Cyclotron Road, Berkeley CA 94720, USA}
\affiliation[h]{New York University Abu Dhabi, PO Box 129188, Abu Dhabi, United Arab Emirates}
\affiliation[i]{Center for Astrophysics and Space Science (CASS), New York University Abu Dhabi}

\emailAdd{zhou.conghao@ucsc.edu}

\abstract{Recent studies suggest that the stars in the outer regions of massive galaxies trace halo mass better than the inner regions and that an annular stellar mass provides a low scatter method of selecting galaxy clusters. However, we can only observe galaxies as projected two-dimensional objects on the sky. In this paper, we use a sample of simulated galaxies to study how well galaxy stellar mass profiles in three dimensions correlate with halo mass, and what effects arise when observationally projecting stellar profiles into two dimensions. We compare 2D and 3D outer stellar mass selections and find that they have similar performance as halo mass proxies and that, surprisingly, a 2D selection sometimes has marginally better performance. We also investigate whether the weak lensing profiles around galaxies selected by 2D outer stellar mass suffer from projection effects. We find that the lensing profiles of samples selected by 2D and 3D definitions are nearly identical, suggesting that the 2D selection does not create a bias. These findings underscore the promise of using outer stellar mass as a tool for identifying galaxy clusters.}

\begin{document}
\maketitle
\flushbottom

\section{Introduction}

Clusters of galaxies form at the highest density peaks of the cosmological density field, which encode rich information about the underlying cosmology \citep{pressFormationGalaxiesClusters1974,kaiserSpatialCorrelationsAbell1984,bardeenStatisticsPeaksGaussian1986}. Across the electromagnetic spectrum, observable tracers of these density peaks -- such as X-ray luminosity \citep{ghirardiniSRGEROSITAAllSky2024}, the thermal Sunyaev-Zeldovich signal \citep{sehgalAtacamaCosmologyTelescope2011,bocquetClusterCosmologyConstraints2019}, and the velocity dispersion of galaxies in clusters \citep{jingVelocityDispersionProfiles1996, borganiCosmologyUsingCluster1997,caldwellCosmologyVelocityDispersion2016} -- have been carefully studied to extract cosmological information \citep{allenCosmologicalParametersObservations2011}. Linking tracers to cosmology requires accurately characterizing their relationship with their host dark matter (DM) halos \citep{geachClusterRichnessMass2017, simetWeakLensingMeasurement2017, melchiorWeaklensingMassCalibration2017, miyatakeWeaklensingMassCalibration2019, mcclintockDarkEnergySurvey2019, schrabbackMassCalibrationDistant2021, robertsonACTDR5SunyaevZeldovichClusters2024, grandisSRGEROSITAAllSky2024}. Since the advent of wide-field optical and near-infrared sky surveys \citep{yorkSloanDigitalSky2000, dejongKiloDegreeSurvey2013, flaugherDarkEnergyCamera2015, aiharaHyperSuprimeCamSSP2018}, weak gravitational lensing \cite{bartelmannWeakGravitationalLensing2001}, the coherent distortion of background source galaxies, has been the top choice for calibrating the relation between tracers and their dark matter halos for two reasons. First, gravitational lensing probes the total matter content and thus eliminates the need for assumptions about the dynamical state of the halo. Second, by grouping the lensing signals of haloes with similar observed properties, stacked weak gravitational lensing can yield very precise constraints on the mean mass of a population of haloes \citep{wuCosmologyGalaxyCluster2021}. These advantages of weak gravitational lensing make optical tracers ideal for constraining cosmology, as the tracers can be identified from the same lensing data.

One extensively studied optical tracer of cluster mass, is the number of red and quenched galaxies in a galaxy cluster, commonly known as \emph{richness} \citep{rykoffRedMaPPerAlgorithmSDSS2014, rozoRedMaPPerIIIDetailed2015}. The current richness-based optical cluster cosmology methodology groups clusters by richness and simultaneously constrains the halo mass-richness relation and cosmological parameters through stacked lensing signals and cluster number counts \citep{costanziMethodsClusterCosmology2019, abbottDarkEnergySurvey2020}. Optical cluster cosmology with richness was very popular \citep{rozoCosmologicalConstraintsSDSS2009, costanziMethodsClusterCosmology2019} for Stage II surveys of the Dark Energy Task Force report \citep{albrechtReportDarkEnergy2006}. However, as data from Stage III optical wide-field surveys continues to lower the statistical uncertainty of the relation between halo mass and richness, important unaddressed and complex systematics (for example, projection effects \citep{mylesSpectroscopicQuantificationProjection2021} and optical selection effects \citep{sunayamaImpactProjectionEffects2020,wuOpticalSelectionBias2022}) have been discovered \citep{abbottDarkEnergySurvey2020}. Direct calibration of these systematics with existing numerical simulations proves challenging because of the difficulty of simulating the satellite distribution in clusters and the selection function of cluster finders. While efforts are underway to accurately model and calibrate the aforementioned systematics \citep{parkClusterCosmologyAnisotropic2021, 
zengSelfcalibratingOpticalGalaxy2023, salcedoDarkEnergySurvey2023, zhouForecastingConstraintsOptical2023, shiIntrinsicAlignmentGalaxy2023, sunayamaOpticalClusterCosmology2023,leeOpticalGalaxyCluster2024}, studies of alternative tracers of the highest-density peaks using optical data are also being explored \citep{pereiraMassesWeaklensingCalibration2020, huangWeakLensingReveals2020, huangOuterStellarMass2021, estevesCopacabanaProbabilisticMembership2024, kwiecienImprovingGalaxyCluster2024}. 

Among the proposed optical tracers, the stellar mass in the outskirts of massive galaxies, also known as outer stellar mass, was found to have a tight correlation with the mass of the host dark matter halo \citep{huangOuterStellarMass2021, kwiecienImprovingGalaxyCluster2024}. Specifically, \cite{huangOuterStellarMass2021} found the stellar mass selected in the annulus between two ellipses with major axes of $50$ and $100$ kpc is a halo mass tracer with a stellar mass to halo mass relation (SHMR) scatter comparable to richness. This was confirmed using the Hyper Suprime-Cam (HSC) \citep{aiharaHyperSuprimeCamSSP2018} and the Dark Energy Survey \citep{abbottDarkEnergySurvey2016} data by \cite{kwiecienImprovingGalaxyCluster2024}. Outer stellar mass offers two main advantages over richness. First, richness-based cluster finders usually use the photometric redshift of galaxies to find cluster members for a large sample of clusters. The large uncertainty of photometric redshifts couples richness with the structures along the line-of-sight and creates projection effects in lensing \citep{parkClusterCosmologyAnisotropic2021, wuOpticalSelectionBias2022,leeOpticalGalaxyCluster2024, xuOutskirtStellarMass2024}. Compared to richness, the outskirt stellar mass of massive central galaxies is not expected to suffer these kinds of projection effects, since we can identify the massive central galaxy first and only use redshifts of the massive galaxies of interest. However, it is possible that the outskirt stellar mass is also prone to the projection effect induced by the triaxial shape of halos \citep{osatoStrongOrientationDependence2018, zhang3DIntrinsicShapes2022}, which couples the orientation of an ellipsoidal halo and the mass proxy. In this work, we will show that the impact of triaxiality on outer stellar mass is not significant. Second, finding a sample of massive galaxies above a stellar mass threshold should be a more straightforward process than finding galaxy clusters with an overdensity of red-sequence galaxies. This means that massive galaxy finder may be easier to calibrate and less prone to complex systematics. These properties give the outskirts stellar mass of massive central galaxies the potential to be a cleaner and more informative link between the observed proxy of high-density peaks and the underlying cosmology. With current and future optical survey data and improvements in low surface brightness photometry, the outskirt stellar mass could be an independent probe with comparable constraining power as richness \citep{xhakajClusterCosmologyCluster2023}. Moreover, outer stellar mass can be combined with richness to probe halo assembly history (Xu et al. in prep.) or to calibrate the mass-richness relation \citep{kwiecienImprovingGalaxyCluster2024}. Building on these promising prospects, \cite{xhakajClusterCosmologyCluster2023} uses numerical simulations to forecast that, by combining clustering, lensing, and outer stellar mass from a stellar mass complete sample from the Dark Energy Spectroscopic Instrument Bright Galaxy Survey \citep{hahnDESIBrightGalaxy2023}, massive galaxies have the potential to yield competitive constraints on $\sigma_8$, the present root-mean-square matter fluctuation
averaged over a sphere of radius $8h^{-1}$ Mpc, and $\Omega_m$, the matter density parameter, with controllable systematics. 

\begin{figure} 
\includegraphics[width=\columnwidth]{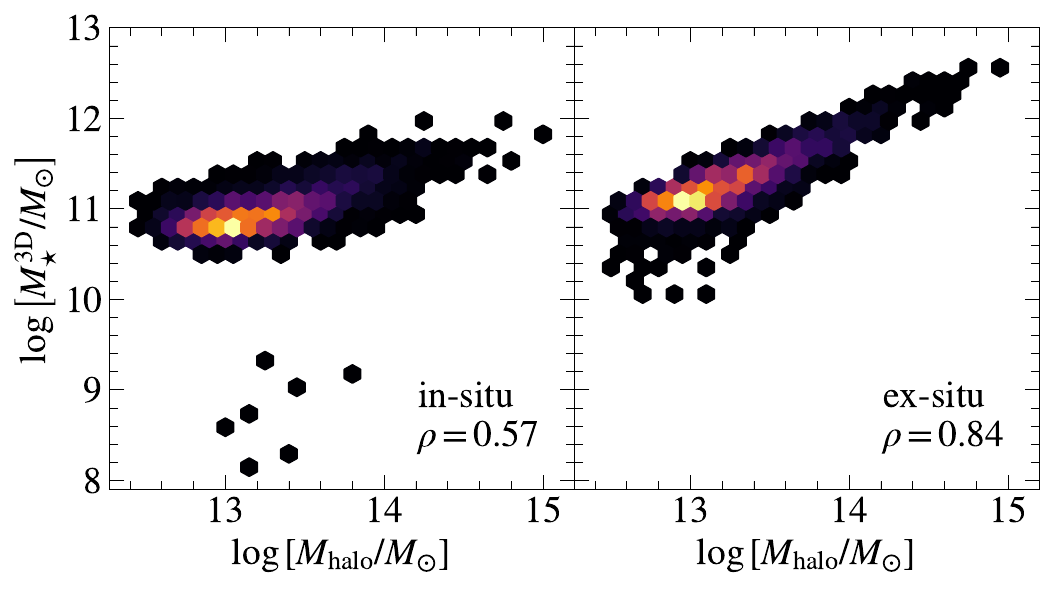}
\caption{Joint distribution of halo mass and central galaxy stellar mass in IllustrisTNG 300-1. On the left, we see that the in-situ stellar mass changes very little as the halo mass changes. On the right, the logarithmic ex-situ stellar mass increases linearly as the logarithmic halo mass increases. The correlation efficient $\rho$ between ex-situ stellar mass and halo mass is much higher than that between in-situ stellar mass and halo mass. More details about the simulation and the labeling of in-situ and ex-situ particles can be found in Section~\ref{sec:simulation}.} 
\label{fig:in_ex_correlation}
\end{figure}

The observation in \cite{huangOuterStellarMass2021} that the stellar mass in the outskirts of massive galaxies correlates well with the halo mass is well-explained by the two-phase model of galaxy formation \citep{oserTwoPhasesGalaxy2010}. In this model, massive galaxies first undergo a phase of rapid star formation where the in-situ stars form from infalling cold gas \citep{dekelColdStreamsEarly2009}. This first phase is followed by a subsequent longer phase during which massive galaxies grow a major fraction of their stellar mass mainly by accreting smaller satellite galaxies \citep{naabMinorMergersSize2009, bezansonRelationCompactQuiescent2009, vandokkumGrowthMassiveGalaxies2010, cooperGalacticStellarHaloes2010, cooperGalacticAccretionOuter2013, rodriguez-gomezStellarMassAssembly2016}. These accretions leave diffuse stellar trails in the outskirts of the central halo \citep{helmiBuildingStellarHalo1999, bullockTracingGalaxyFormation2005, amoriscoContributionsAccretedStellar2016}, and the long dynamical time in the outskirts preserves the trails throughout the life of the massive galaxy \citep{helmiStellarHaloGalaxy2008, cooperGalacticStellarHaloes2010}. The stars of a low-redshift massive central galaxy can thus be partitioned into an in-situ stellar component that has little correlation with the galaxy assembly and an ex-situ stellar component that is a fossil record of the mass assembly history of the galaxy \citep{rodriguez-gomezStellarMassAssembly2016, pillepichFirstResultsIllustrisTNG2018}. The close connection between accreted stars and halo assembly history has been studied extensively in numerical simulation \citep{pillepichFirstResultsIllustrisTNG2018, montenegro-tabordaGrowthBrightestCluster2023, ahvaziProgenitorsIntraclusterLight2023,brownAssemblyIntraclusterLight2024}, semi-analytical models \citep{deluciaHierarchicalFormationBrightest2007}, and in observations \citep{shankarAvoidingProgenitorBias2015, golden-marxCharacterizingIntraclusterLight2023, golden-marxHierarchicalGrowthBright2024, zhangDarkEnergySurvey2024}. Given the close connection between the ex-situ stellar component and the halo assembly history, the ex-situ component %
should be more correlated with the overall halo mass compared to the in-situ component. We see that this is indeed the case from a quick inspection of a hydrodynamical simulation. In the left (right) panel of Figure~\ref{fig:in_ex_correlation}, we show the joint distribution of halo mass and total in-situ (ex-situ) stellar mass of massive galaxies in IllustrisTNG 300-1 \citep{nelsonIllustrisTNGSimulationsPublic2019} (See Section~\ref{sec:simulation} for more details). We see that the ex-situ component changes with halo mass as a power law. In contrast, the in-situ component changes very little as the halo mass changes. The ex-situ stellar mass is much better correlated with the overall halo mass, especially in the low mass regime.

Given that ex-situ stars correlate better with halo mass than in-situ stars, for a selection of stars in a galaxy, a higher ex-situ fraction should correspond to a better correlation with halo mass. In simulations, it is observed that the ex-situ fraction is higher in 3D spherical shells in the outskirts compared to spherical shells near the galaxy core \citep{rodriguez-gomezStellarMassAssembly2016, pillepichFirstResultsIllustrisTNG2018}. However, in observations, we only see the 2D projection of the stellar profile of the massive galaxies, and thus, we are limited to selecting stars based on their projected profiles. Combined with the fact that massive central galaxies are usually highly non-spherical \cite{liOriginPropertiesMassive2018} and thus susceptible to projection effects, the limitation of 2D selection poses questions about the efficacy of the outer stellar mass selection and its relation to the ex-situ stellar population. Do projection effects impact the selection of high ex-situ parts in the galaxy? Do projection effects create unexpected systematic effects that couple 2D outer stellar mass with lensing, which could potentially bias cosmology constraints if left unmodeled?

In this paper, we try to answer these questions through a more general approach. We study why the stellar mass from some 2D selections performs better than others. We also try to identify an optimal way of defining a tracer based on a 2D selection of stars that best correlates with halo mass. We use massive galaxies from the state-of-the-art numerical simulation IllustrisTNG, which is introduced in Section~\ref{sec:simulation}. In Section~\ref{sec:methods}, we introduce the methods used to define and evaluate tracers as a halo mass proxy based on 2D and 3D selections. In Section~\ref{sec:results}, we demonstrate and explain that the outskirt stellar mass selected by a 2D elliptical annulus is the optimal halo mass tracer for selections defined in 2D. We also show that there is a simple mapping between the 2D elliptical annulus outer stellar mass and its 3D counterpart with little scatter, and there are minimal differences between lensing profiles around massive central galaxies selected by 2D and 3D outer stellar mass. Finally, we conclude by discussing some potential caveats and summarize our main findings in Section~\ref{sec:summary}.

\section{Numerical simulation and sample selection}
\label{sec:simulation}
In this section, we present the numerical simulation and the derived data products used to analyze the stellar content of massive galaxies. We first introduce the IllustrisTNG simulations and the definition of our galaxy sample. Then, we briefly explain the distinction between in-situ and ex-situ stellar particles in the simulation. Finally, we describe the creation of the 2D projected stellar maps from which we extract the 2D stellar mass.

\subsection{The IllustrisTNG Simulations}
 IllustrisTNG is a suite of $\Lambda$CDM magnetohydrodynamic simulations that enables studies of galaxies and galaxy clusters \citep{pillepichSimulatingGalaxyFormation2018,nelsonIllustrisTNGSimulationsPublic2019} in a cosmological setting. IllustrisTNG uses the tree-particle-mesh code \textsc{arepo} \citep{springelPurSiMuove2010,sijackiMovingMeshCosmology2012, pakmorSimulationsMagneticFields2013, pakmorImprovingConvergenceProperties2016, weinbergerAREPOPublicCode2020} to simulate large scale structure and galaxy formation in three volumes with side lengths of $\sim 50 \; \mathrm{Mpc}$, $\sim 100 \; \mathrm{Mpc}$, and $\sim 300 \; \mathrm{Mpc}$ \citep{pillepichSimulatingGalaxyFormation2018}. The galaxy formation model includes radiative cooling and heating, star formation, stellar evolution, galactic winds, and the formation, growth, and energetic feedback of supermassive black holes \citep{weinbergerSimulatingGalaxyFormation2017,pillepichSimulatingGalaxyFormation2018, naimanFirstResultsIllustrisTNG2018, nelsonFirstResultsIllustrisTNG2018}. The simulated stellar observables, including the galaxy stellar mass function and stellar-to-halo mass relation (SHMR) \cite{pillepichFirstResultsIllustrisTNG2018}, galaxy color bimodality \cite{nelsonFirstResultsIllustrisTNG2018}, the clustering of galaxies as a function of stellar mass \citep{springelFirstResultsIllustrisTNG2018}, and massive galaxy stellar density profiles \cite{ardilaStellarWeakLensing2021}, are in good agreement with observational data. Due to resolution effects, the stellar mass between different TNG boxes does not agree perfectly \citep{pillepichFirstResultsIllustrisTNG2018}. For example, the stellar mass of a halo in TNG100 is $\sim$ 1.4 times higher than a similar halo in TNG300 \citep{pillepichFirstResultsIllustrisTNG2018}. We do not explicitly correct for this effect since we are not comparing our results from different boxes with different resolutions or comparing our results to data. If we adopt the fix of multiplying the stellar mass in TNG300 by 1.4 \cite{pillepichFirstResultsIllustrisTNG2018}, the aforementioned resolution effects only change the overall normalization of stellar mass, and thus, it should not impact the scatter of the SHMR which is one of the key aspects studied here.

 To reduce sample variance for our massive cluster sample without compromising our science goals, in this paper, we use the highest mass resolution version of the largest volume boxes in the suite, TNG300-1. For TNG300-1, the DM (baryon) mass resolution is $6.3 \times 10^{6} \Msun$ ($1.3 \times 10^6 \Msun$) and the side length is 302.6 Mpc. The minimal physical scale we study in this paper, 30 kpc, is much larger than the 1.4 kpc softening length of TNG300-1 \citep{pillepichFirstResultsIllustrisTNG2018}. In this work, we focus on the most massive galaxies by selecting central galaxies with stellar mass greater than $10^{11.2}M_{\odot}$ at $z = 0.4$. Out of the 2713 selected galaxies, the least massive galaxy has $\sim 2 \times 10^4$ stellar particles. The cosmology parameters are adopted from \cite{adePlanck2015Results2016} with $\Omega_m=0.3089, \Omega_\Lambda = 0.6911, \Omega_b=0.0486, h=0.6774, \sigma_8=0.8159,$ and $n_s=0.9667$.

Halos in IllustrisTNG simulations are identified using the Friends-of-Friends (FoF) algorithm with a linking length of 0.2 times the mean inter-particle distance \citep{davisEvolutionLargescaleStructure1985}. Subhalos are identified as particles bound to overdensities in halos with \textsc{subfind} \citep{springelPopulatingClusterGalaxies2001,dolagSubstructuresHydrodynamicalCluster2009}. In this paper, we use the total mass of the FoF group as the halo mass. We define the central subhalo identified by \textsc{subfind} as the central galaxy \citep{springelPopulatingClusterGalaxies2001, dolagSubstructuresHydrodynamicalCluster2009}.

\subsection{Definition of in-situ and ex-situ components}

In this work, we use the in-situ and ex-situ labels from the stellar assembly catalog \citep{rodriguez-gomezStellarMassAssembly2016} built from the \textsc{sublink} baryonic merger tree \citep{rodriguez-gomezMergerRateGalaxies2015}. The merger tree is constructed by first assigning each subhalo a unique descendant in the next snapshot by comparing shared particles weighted by the bounding energy. The main progenitor is defined as the most massive progenitor halo of a descendent. In this way, a unique main progenitor branch can be constructed by linking all first progenitors. A stellar particle is labeled in-situ if the galaxy in which it formed is in the main progenitor branch. Other stellar particles are labeled as ex-situ. 

\subsection{Stellar Map Making}
We use the analysis code \hydrotools \cite{diemerCOLOSSUSPythonToolkit2018, diemerAtomicMolecularGas2019} to produce projected stellar maps for our sample of massive central galaxies with stellar mass greater than $10^{11.2} \Msun$. Starting from the center of a massive galaxy, we select all stellar particles in a 300 kpc cube that are in the FoF group but do not belong to a satellite galaxy. This is different from how \textsc{subfind} assigns particle membership. The gravitational binding criterion used by \textsc{subfind} includes virtually all stellar particles in the galaxy, but the sample of particles selected by this criterion becomes incomplete at the outskirts of stellar halos compared to the observational data (Leidig et al. in prep.). Selecting all particles not bound to satellite galaxies avoids underestimating the galaxy content in the outskirts (Leidig et al. in prep.), which is critical to assessing the correlation between halo mass and outskirt stellar mass. We then project all particles selected along the $z$-axis to produce $300$ kpc $\times$ $300$ kpc projected stellar maps with 300 pixels on each side. In this work, we focus on the intrinsic relation between the 3D and projected 2D properties of galaxies rather than observational systematics. Therefore, we use the 2D stellar maps directly and do not consider observational effects such as the conversion of magnitude to mass, the impact of the point spread function, and inaccurate background subtraction.

\section{Methods}
\label{sec:methods}
In this section, we introduce the methods used to produce mock observables and quantify the performance of various stellar mass definitions as halo mass proxies. We first introduce the definition of 2D and 3D outskirt stellar masses. We then describe how we fit an SHMR to a given sample of massive galaxies and how we use the best-fit parameters to quantify the performance of each stellar mass selection. Finally, we present the method used to produce mock galaxy-galaxy lensing observables. 

\subsection{2D Isophote Fitting and Elliptical Annulus Stellar Mass}
\label{sec:2d_selection}
To compute 2D elliptical annulus stellar masses, we follow the methodology of \cite{xuOutskirtStellarMass2024} (also see \cite{ardilaStellarWeakLensing2021, cannarozzoContributionSituEx2023}), which closely follows the procedure of extracting 2D elliptical annulus stellar mass in data from \citep{huangOuterStellarMass2021}. In this section, we briefly outline the procedure and refer to \cite{xuOutskirtStellarMass2024} for more details. Given a mock stellar map and a fixed major axis length, we use the isophotal fitting method from \cite{jedrzejewskiCCDSurfacePhotometry1987} implemented in \photutils \citep{bradleyPhotutilsPhotometryTools2016} to extract elliptical isophotes from stellar maps. Specifically, we use steps of 10\% of the current ellipse semi-major axis length to grow or shrink the semi-major axis. We use 2$\sigma$ clipping three times to clip out outliers pixels. We always fix the center of the ellipse to the center of the projected galaxy. After obtaining isophotes with different semi-major axis lengths, we calculate the flux-weighted mean ellipticity and position angle. These are then used to define an elliptical annulus given an inner and outer semi-major axis length. In this paper, the integrated stellar mass in the annulus is referred to as the elliptical annulus stellar mass or 2D outer stellar mass. 

\subsection{3D Ellipsoid Estimation and Ellipsoidal Shell Stellar Mass}
\label{sec:ellipsoid}
Our method for computing the 3D ellipsoidal shell selection follows a similar procedure as the 2D elliptical annulus selection. Given a set of stellar particles in a galaxy, we first estimate the direction of the principal axes and the axes ratios following the method in \cite{dubinskiStructureColdDark1991}. We then use the best-fit principal axes directions and axis ratios to find the mass of stellar particles in a shell bounded by 3D ellipsoids with an inner and outer semi-major axis length. Specifically, we estimate the direction of the principal axes and axis ratios by iteratively calculating the reduced inertia tensor. The ellipsoidal radius is defined as

\begin{equation} \label{eq:elip_r}
    r =\left(x_1^2+\frac{x_2^2}{q^2}+\frac{x_3^2}{s^2}\right)^{1 / 2},
\end{equation} 

\noindent where $x_1, x_2, x_3$ are the coordinate components of a particle in the principal axis frame of an ellipsoid, while $q$ and $s$ are the intermediate-to-major and minor-to-major axis ratios. The ellipsoidal radius of a particle is the semi-major axis length of the corresponding 3D isophote.

Starting with $q=s=1$ in the simulation coordinate frame, we iteratively calculate the reduced inertia tensor

\begin{equation}
    \mathcal{M}_{i j}=\sum_{p=1}^N m_p \frac{r_i r_j}{r^2}.
\end{equation}

The summation iterates over all stellar particles, $r_i$ and $r_j$ stand for the $i$-th and $j$-th coordinate component in the current frame, and $m_p$ is the mass of the stellar particle. By diagonalizing the reduced moment of inertia tensor, we obtain the axis ratios $q,s$ and the directions of the principal axes. We then rotate the coordinate frame into the principal axes frame and recalculate the reduced moment of inertia tensor in this new frame. The algorithm repeats until the change of the axis ratios between two iterations is smaller than $10^{-8}$. We take the axis ratios and the principal axis directions in the last iteration as the best-fit values. We refer the readers to Appendix~\ref{appendix:moment} for more details.

With the best-fit axis ratios and the direction of the principal axes, at any given semi-major axis length, we can obtain an ellipsoid with a similar shape as the galaxy. Given an inner and outer semi-major axis length, we sum all stellar particle masses between the ellipsoids defined by the inner and outer major axis lengths to get an ellipsoidal shell stellar mass or 3D outer stellar mass.

With the best-fit axis ratios $q$ and $s$ and the direction of the principal axes, we can also calculate the ellipsoidal radius for any particle with Eq.~\ref{eq:elip_r} in a given galaxy. In the rest of the paper, we use ellipsoidal polar coordinates parameterized by the ellipsoidal radius $r$ and the angle between the particle and the major axis of the galaxy $\theta$ to describe particles' relative 3D positions in galaxies. This coordinate system is convenient since it enables us to compare quantities at the same relative position in galaxies with different shapes.

\subsection{Stellar-to-Halo Mass Relation}
\label{sec:shmr}
We quantify the performance of various stellar mass selections as halo mass proxies using the scatter of the SHMR. We model the relation between stellar mass $M_*$ and halo mass $M_h$ with a log-linear relation with a constant Gaussian scatter. We assume that the conditional distribution of halo mass at a given stellar mass follows a Gaussian distribution

\begin{equation}
   \log_{10}(M_h) | \log_{10}(M_*) \sim \mathcal{N} \left( \alpha \log_{10}(M_*) + \beta , \sigma_{M_h | M_*}^2 \right),
\end{equation}

\noindent where $\alpha$ and $\beta$ are the slope and intercept of the log-linear relation and $\sigma_{M_h |M_*}$ is the scatter of halo mass at fixed stellar mass. Throughout the remainder of the paper, we refer to this as the ``scatter of the SHMR"\footnote{The scatter in stellar mass at fixed halo mass is another way of measuring scatter in the SHMR.}. In general, we'd like to use definitions of $M_*$ that minimize the scatter in the SHMR. Also, we'd like definitions of $M_*$ with shallow slopes ($\alpha$) since the observational uncertainty in stellar mass $\sigma_{M_*}$ scales to the uncertainty in halo mass as $\alpha \times \sigma_{M_*}$.

Given the scatter of outer stellar mass at a given total stellar mass, a total stellar mass complete sample is not necessarily complete in outer stellar mass. For example, if the smallest outer stellar mass is $\mu$ in our sample of central galaxies with a stellar mass cut of $10^{11.2} \Msun$, the sample is not necessarily complete in outer stellar mass above $\mu$. There may be some galaxies with outer stellar mass greater than $\mu$ but total stellar mass smaller than $10^{11.2} \Msun$ that are excluded in our sample. To alleviate the impact of incompleteness in outer stellar mass induced by our cut in total stellar mass, we follow the method in \cite{xuOutskirtStellarMass2024} and only use galaxies with $\log[M_*] > \mpeak$ in which $\mpeak$ is the value of $M_*$ where the stellar mass function peaks. 

\begin{figure}
\includegraphics[width=0.6\columnwidth]{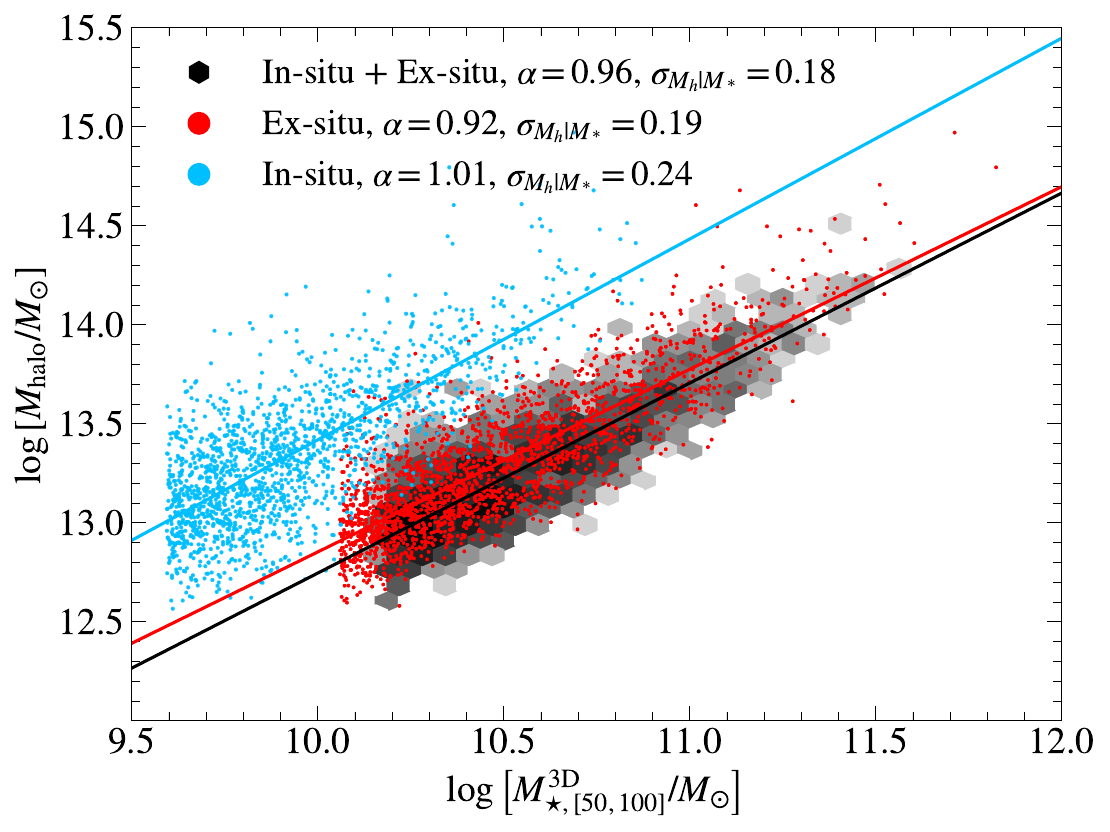}
\centering
\caption{SHMR for the in-situ and ex-situ mass located within a 3D ellipsoidal shell of $[50,100]$ kpc. The red (cyan) points represent ex-situ (in-situ) stellar mass in $[50,100]$ kpc ellipsoidal shells. The cyan (red) lines are the best fit SHMR described in Section~\ref{sec:shmr}. The black hexagons are the total stellar mass (in-situ + ex-situ) in $\textrm{3D}[50,100]$ kpc ellipsoidal shells. The SHMR of the in-situ component has a larger scatter and a larger slope compared to the SHMR of the ex-situ component. Thus, the ex-situ component is a better tracer of halo mass than the in-situ component.}
\label{fig:eg_shmr}
\end{figure}

Fig.~\ref{fig:eg_shmr} shows an example fit to the SHMR between 3D ellipsoidal shell stellar mass with inner and outer radii of $[50, 100]$ kpc. The sharp vertical edges on the low mass end are from the completeness cut. We see that the in-situ component has a larger scatter and slope compared to the ex-situ component. Therefore, the ex-situ component is a better tracer of halo mass (this will be quantified further in Section~\ref{sec:shmr_scatter}).

\subsection{Weak lensing observable}

Galaxy-galaxy lensing is commonly used to infer the halo masses of an ensemble of galaxy clusters \citep{melchiorWeaklensingMassCalibration2017, simetWeakLensingMeasurement2017, murataConstraintsMassRichnessRelation2018, mcclintockDarkEnergySurvey2019}. Results from \cite{huangOuterStellarMass2021} suggest that outer stellar mass traces halo mass as well as, and perhaps even better than, richness. However, a key question of concern is whether or not an outer stellar mass selection might lead to projection effects that impact the lensing profiles. Here, we compare lensing profiles for total mass selections with 2D and 3D selections and study if there are any scale-dependent effects in the resulting lensing profiles. For this, we need to calculate the lensing profile around clusters in TNG300. We use \textsc{halotools} \citep{hearinForwardModelingLargescale2017} to calculate the mock weak lensing observable $\dsigma(\rp)$, the excess surface density around the halo center. Physically, $\dsigma(\rp)$ is defined as
\begin{equation}
    \dsigma(\rp) = \widebar{\Sigma}(< \rp) - \Sigma(\rp) , 
\end{equation}
where $\Sigma(\rp)$ is the azimuthally averaged surface density with projected distance $\rp$ from the halo center, and $\widebar{\Sigma}(<\rp)$ is the mean surface density in a disk with radius $\rp$ around the halo center defined as 

\begin{equation}
\widebar{\Sigma}(\rp) = \frac{1}{\pi \rp^2} \int_0^{\rp} \Sigma(\rp') \; 2 \pi \rp' \mathrm{d} \rp' .
\end{equation}

In practice, given the particle positions in a simulation,  calculates the two-point correlation function between the position of the halo center and the position of DM, gas, and stellar particles $\xi_{i}(\rp, z)$, in which $\rp$ is the projected distance from the halo center and $z$ is the line of sight distance. The two-point correlation function is related to the surface density profile with 

\begin{equation}
  \Sigma(\rp) =\int_0^L \; \sum_i m_i \; \xi_i\left(\rp, z \right) \; \mathrm{d} z,
\end{equation}
where $m_{i}$ is the mass of the particle $i$ and $L$ is the box length of the simulation.

\subsection{Comparing 2D and 3D quantities}
To compare 2D and 3D stellar masses, we need to take into account the projection of 3D length scales. To find the mapping between the length of a projected 2D vector and its 3D counterpart, we define the conversion factor from 3D length to 2D length $A$ as follows:

\begin{equation}
    A = \frac{2}{\pi} \int_0^{\frac{\pi}{2}} \cos \theta \; d \mathrm{\theta} \approx 0.63,
\end{equation}
which corresponds to the mean length of a randomly projected 3D unit vector. In the remainder of this paper, when there is a comparison between 2D and 3D quantities, we use the following equation to map 2D quantities to their 3D counterparts, 

\begin{equation}
    l_{\mathrm{3D}} =  \frac{l_{\mathrm{2D}}}{A},
    \label{eq:2to3}
\end{equation}
in which $l_{\mathrm{2D}}$ is any 2D length and $l_{\mathrm{3D}}$ is the 3D counterpart of the 2D length. Given this conversion, assuming random LOS, the mean length of the semi-major axis of the projected ellipse of a highly prolate ellipsoid with semi-major axis $l_{\mathrm{3D}}$ is $l_{\mathrm{3D}} \times A$. We also note explicitly if a length is a 3D or projected 2D quantity with superscripts. For example, $\solida$ is the 3D semi-major axis length of an ellipsoid and $\planea$ is the 2D semi-major axis length of an ellipse. 

\section{Results}
\label{sec:results}
In this section, we first analyze the 3D shape distribution of massive central galaxies in IllustrisTNG. We also show that the projected 2D shape can vary significantly depending on lines of sight (LOS). We study the spatial distributions of in-situ and ex-situ mass and argue that ellipsoidal shells are the optimal way to select stellar particles because they maximize the ex-situ fraction in 3D. We show that stellar mass defined by a 2D elliptical annulus performs comparably or even better than that selected by 3D ellipsoidal shells as a halo mass proxy. We explain this counterintuitive result with a simple and intuitive model. Finally, we show that lensing profiles of 2D outer stellar mass-selected sample are close to those of corresponding 3D outer stellar mass-selected samples.

\subsection{The shapes of massive galaxies in TNG300}
\label{sec:morphology}
Assuming that galaxy shapes can be approximated with ellipsoids, we can fully describe the morphologies of massive galaxies with the three parameters $\solidb/\solida, \solidc/\solida, \solida$. Here $\solidb/\solida$ is the intermediate-to-major axis ratio, $\solidc/\solida$ is the minor-to-major axis ratio, and $\solida$ is the semi-major axis length. Among the three parameters, $\solidb/\solida$ and $\solidc/\solida$ describe the shape of a galaxy, and $\solida$ describes the size of the galaxy. Galaxy shapes can also be classified into discrete categories following different classification schemes \citep{jiangFormationUltradiffuseGalaxies2019}. In this paper, we follow the classification scheme of \cite{liOriginPropertiesMassive2018} to classify galaxies in TNG300 at $z=0.4$ with $\mtot >10^{11.2} \Msun $ into these four categories:

\begin{equation}
    \text{A galaxy is} \; 
    \begin{cases}
    \text{Spherical} & \text{if}\; \frac{\solidb}{\solida} - \frac{\solidc}{\solida} \geq 0.2 \text{ and } \frac{\solidb}{\solida} > 0.8\\
    \text{Prolate}   & \text{if}\; \frac{\solidb}{\solida} - \frac{\solidc}{\solida} \geq 0.2 \text{ and } \frac{\solidb}{\solida} \leq 0.8  \\
    \text{Triaxial}  & \text{if}\; \frac{\solidb}{\solida} - \frac{\solidc}{\solida} < 0.2 \text{ and } \frac{\solidb}{\solida} \leq 0.8 \\
    \text{Oblate}    & \text{if}\; \frac{\solidb}{\solida} - \frac{\solidc}{\solida} < 0.2 \text{ and } \frac{\solidb}{\solida} > 0.8\\
    \end{cases}.
\end{equation}

\begin{figure}[h]
\centering
\includegraphics[width=0.6\columnwidth]{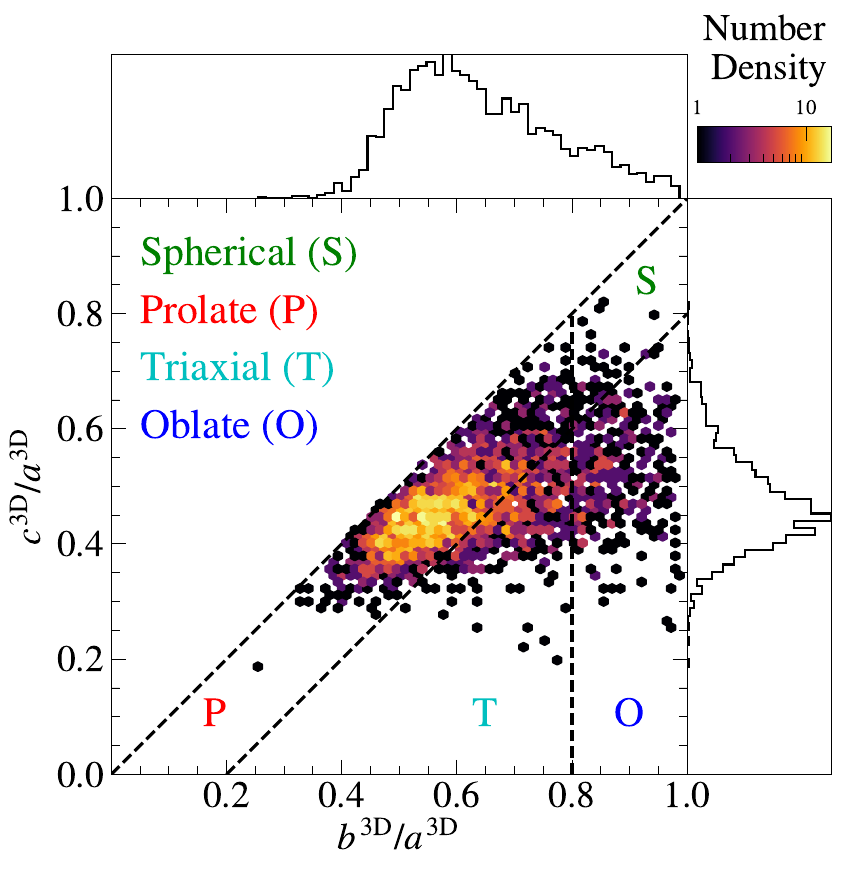}
\caption{Axis ratio distribution of massive central galaxies in TNG300. The axis ratio distribution peaks around $(0.5,0.5)$ in the $(\solidb/\solida, \solidc/\solida)$ plane, and there is a strong correlation between $\solidb/\solida$ and $\solidc/\solida$. The majority $\sim 67 \%$ of massive galaxies in our sample fall into the prolate category. 99 \% of the galaxies are non-spherical and thus susceptible to projection effects.}
\label{fig:baca}
\end{figure}

Figure~\ref{fig:baca} shows the classification scheme from \cite{liOriginPropertiesMassive2018} in the $(\solidb/\solida, \solidc/\solida)$ plane. In the same figure, we show the axis ratio distribution of massive galaxies from TNG300 calculated with stellar particles in the $(\solidb/\solida, \solidc/\solida)$ plane. We see that the axis ratio peaks around $(0.5,0.5)$, and there is a strong correlation between $\solidb/\solida$ and $\solidc/\solida$. In this classification scheme,  67\% of galaxies with $\mtot \geq 10^{11.2} M_{\odot}$ are prolate, 18\% are triaxial, and 13\% are oblate. Given that non-spherical galaxies suffer from projection effects, the relative morphology fraction shows that only 1\% of massive galaxies are not impacted by projection.

\begin{figure*}
\includegraphics[width=\columnwidth]{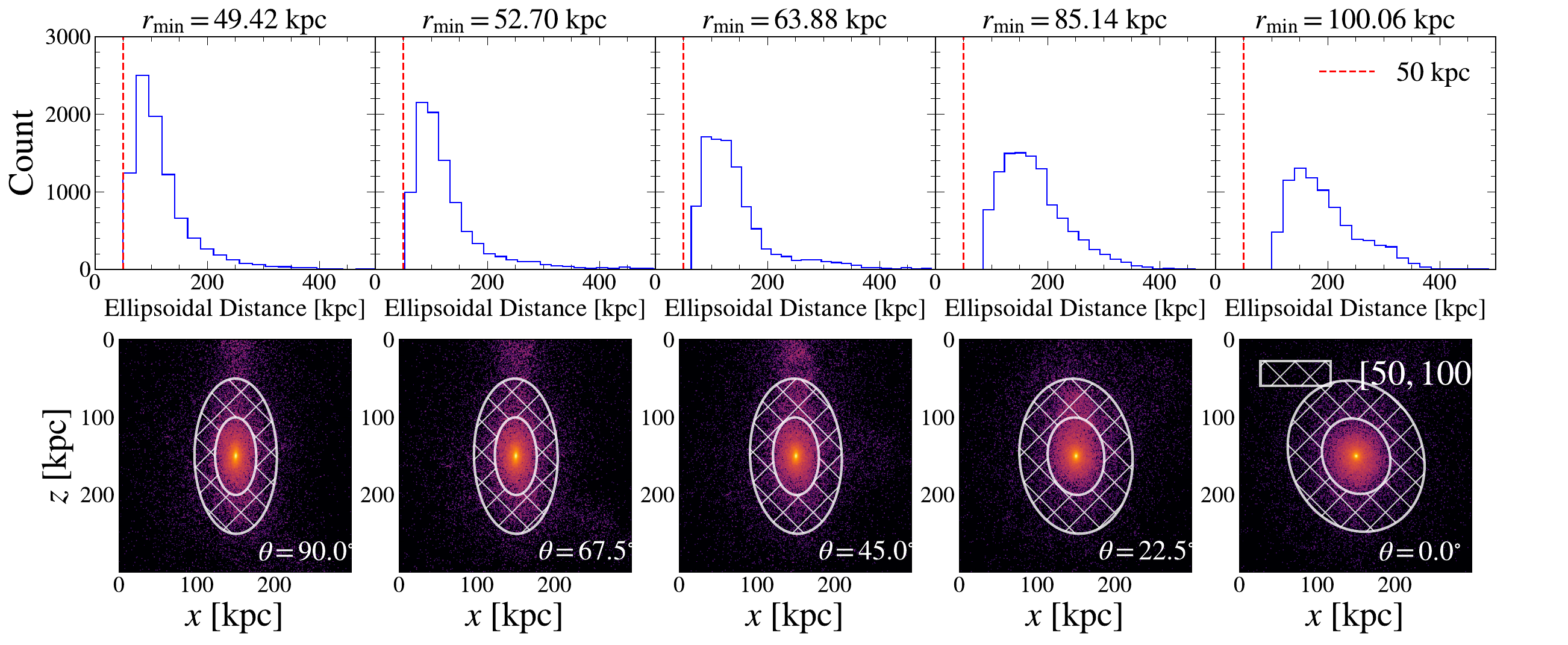}
\caption{Example of how the 2D shape of a prolate galaxy depends on the angle of projection. \textbf{Bottom}: 2D images of the same prolate galaxy viewed from different angles $\theta$ from the major axis. The area covered by the white hatch is the 2D $[50, 100]$ kpc stellar mass definition. From left to right, we show the same prolate galaxy projected at different angles from the major axis. The leftmost panel shows the galaxy projected along its minor axis, and it appears elliptical in 2D. The rightmost panel shows the galaxy projected along its major axis, and it appears more circular in 2D. \textbf{Top}: distribution of 3D ellipsoidal distances of stellar particles selected in a 2D $[50,100]$ kpc annulus. The minimal corresponding ellipsoidal distance is printed at the top of each sub-panel. We see that the same $[50, 100]$ kpc annulus selects populations with different 3D ellipsoidal $r$ distributions. The minimum of the distribution is always greater or approximately equal to 50 kpc (the lower bound of the 2D cut).}
\label{fig:prolate_example}
\end{figure*}

Viewed in 2D, the shape of a galaxy depends both on its intrinsic 3D morphology and its orientation. As an example, the bottom row of Figure~\ref{fig:prolate_example} shows the 2D projection of a prolate galaxy with different viewing angles. We see that the projected shape can either appear as an ellipse or a circle depending on the angle of projection. The leftmost figure is the projection of the prolate galaxy along its minor axis, and the rightmost figure is the projection along its major axis. The top row of Figure~\ref{fig:prolate_example} shows the distribution of 3D ellipsoidal radii of particles selected by the 2D $[50, 100]$ kpc annulus. We see that a 2D selection corresponds to different 3D selections depending on the viewing angle. In the following sections, we will study the in-situ and ex-situ distributions in 2D and quantify the impact of 2D selections.

\subsection{Spatial distribution of ex-situ stellar particles}
\label{sec:ex_situ_distribution}

\begin{figure*}
\includegraphics[width=\textwidth]{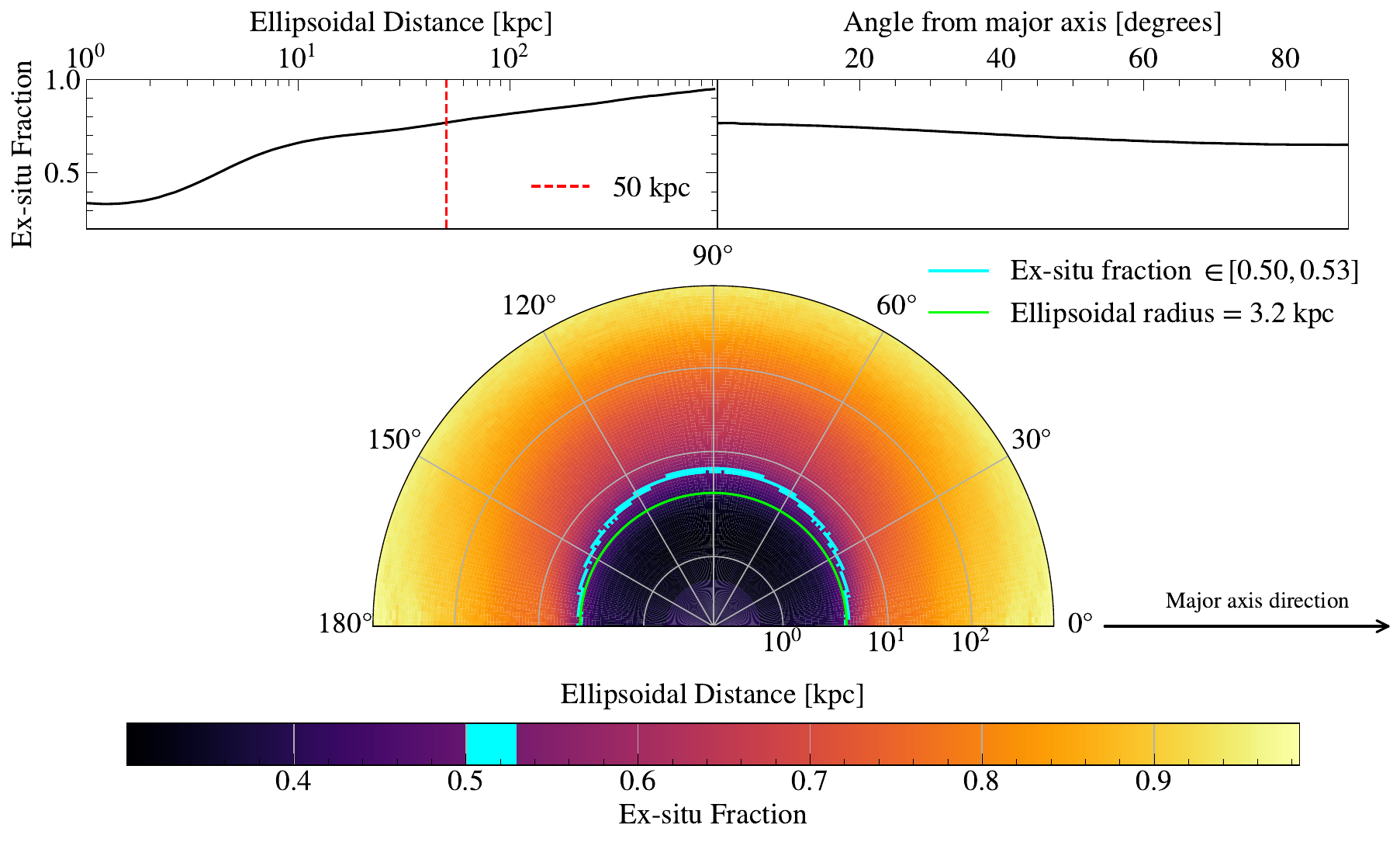}
\caption{The spatial distribution of the ex-situ fraction in polar ellipsoidal coordinates. \textbf{Top}: mean ex-situ fraction as a function of ellipsoidal radius and angle from the major axis. \textbf{Bottom}: mean ex-situ fraction in elliptical polar coordinates. The major axes of the galaxies in this figure are aligned with $0^{\circ}$. The cyan line indicates the region where the ex-situ fraction is between 0.50 and 0.53. The green line is the constant 3.2 kpc ellipsoidal radius line. The ex-situ fraction monotonically increases from the center to the outskirts. In the direction of the major axis of the massive central galaxy, the mean ex-situ fraction is higher than in other directions. This can be seen by comparing the blue constant ex-situ fraction line and the green constant radius line: if the ex-situ fraction had no angular dependence, the cyan and green curves would overlap. However, at a given ellipsoidal radius, the locations near the major axis have a higher ex-situ fraction. Given the radial and angular trends in ex-situ fraction, selecting stars with an ellipsoidal shell that follows the shape of the galaxy and only includes the outskirt stellar masses can maximize the ex-situ fraction in the selection and result in the minimal SHMR scatter.}
\label{fig:ex_situ_frac}
\end{figure*}

Previous work suggests that ex-situ stars correlate better with halo mass than in-situ stars \citep{bradshawPhysicalCorrelationsScatter2020,huangOuterStellarMass2021}. But what is the best approach to measuring these ex-situ stars while also excluding the in-situ component? Here we study the spatial distribution of ex-situ stars to create a more optimized selection. The top row of  Figure~\ref{fig:ex_situ_frac} shows the spatial distribution of the ex-situ fraction in ellipsoidal radii and the angle between particles and the major axis of the galaxy. The top left panel shows that the outskirts of galaxies have a higher ex-situ fraction compared to inner regions. The top right panel shows that there is an anisotropy in the relative fraction of ex-situ stellar mass: the ex-situ fraction is higher in the direction of the major axis of the galaxy compared to the direction of the minor and intermediate axes. The bottom panel of the same figure shows the ex-situ fraction of massive galaxies in an ellipsoidal polar coordinate system defined in Section~\ref{sec:ellipsoid}. The joint distribution exhibits the same and independent trends of higher ex-situ fraction in the outskirts and along the major axis. In the bottom panel, a cyan constant ex-situ fraction curve and a green constant ellipsoidal radius curve are shown to visually aid the illustration. If there is no angular trend in the ex-situ fraction, the cyan constant ex-situ fraction curve should overlap with the green constant ellipsoidal radius curve.

\begin{figure*}
\includegraphics[width=\textwidth]{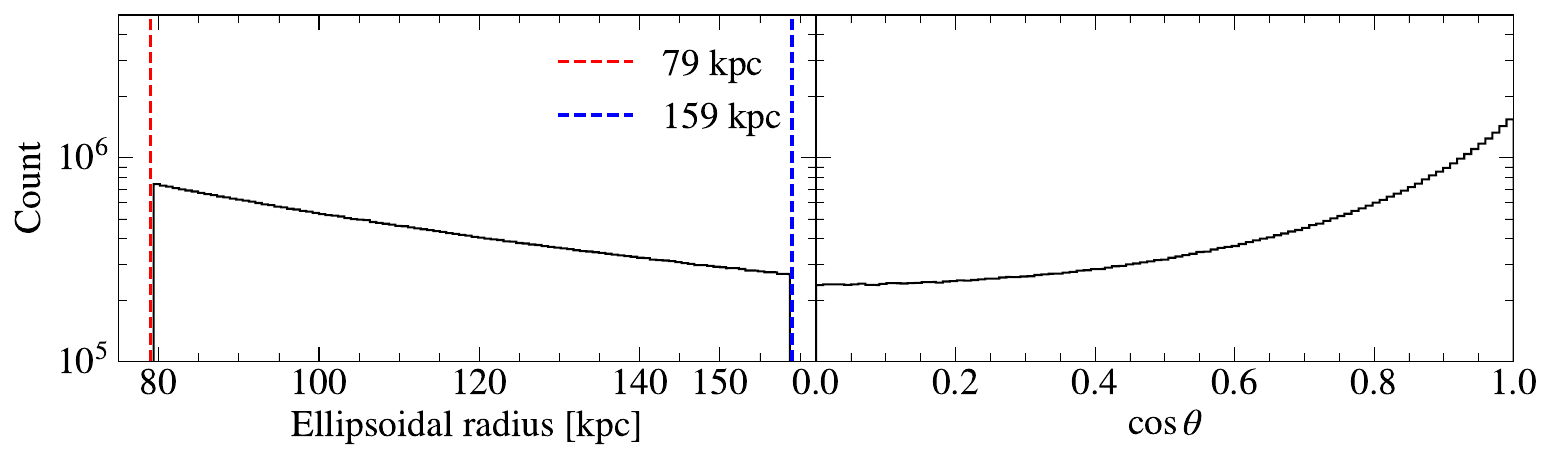}
\caption{Spatial distribution of particles selected with 3D ellipsoidal shells in ellipsoidal polar coordinate system. \textbf{Left:} distribution of ellipsoidal radius of particles selected by 3D ellipsoidal shells with inner radii of 79 kpc and outer radii of 159 kpc, which correspond to 50 kpc and 100 kpc in 2D. \textbf{Right:} distribution of $\cos \theta$ of angle $\theta$ between the major axis of the galaxy and the particles selected by the same ellipsoidal shells. The 3D ellipsoidal shell selection is effective in eliminating the inner part of the galaxy, in this case, particles with an ellipsoidal radius smaller than 79 kpcs. The 3D ellipsoidal selection also preferentially includes particles along the major axis of the galaxy with $\cos \theta$ close to 1. }
\label{fig:3d_radius_angle}
\end{figure*}

Given the angular and radial trends of the ex-situ fraction, we infer that ellipsoidal shells that have similar shapes to the shape of the galaxy light are the best shape for selecting stellar particles that maximize the ex-situ fraction for two reasons. First, ellipsoidal shells select stellar particles that are in the outskirts of the galaxy, which have a higher ex-situ fraction. Second, ellipsoidal shells include more ex-situ components compared to spherical shells because of the anisotropic spatial distribution of the ex-situ stellar particles. In Figure~\ref{fig:3d_radius_angle}, we show the distribution of ellipsoidal radius and the cosine of the angle $\theta$ between each particle and the galaxy's major axis, for stellar particles selected within ellipsoidal shells spanning 79–159 kpc\footnote{79 kpc and 159 kpc corresponds to the 2D 50 kpc and 100 kpc following the conversion in Section~\ref{sec:ellipsoid}.}. The ellipsoidal shells are indeed effective in excluding the inner part of the galaxies, as there are no particles with ellipsoidal radii smaller than 79 kpc in the selection. Also, the ellipsoidal shells preferentially include more particles along the major axis of the galaxy, as shown by the higher density of particles along the major axis with $\cos \theta$ close to 1.

\begin{figure*} [h]
\includegraphics[width=\textwidth]{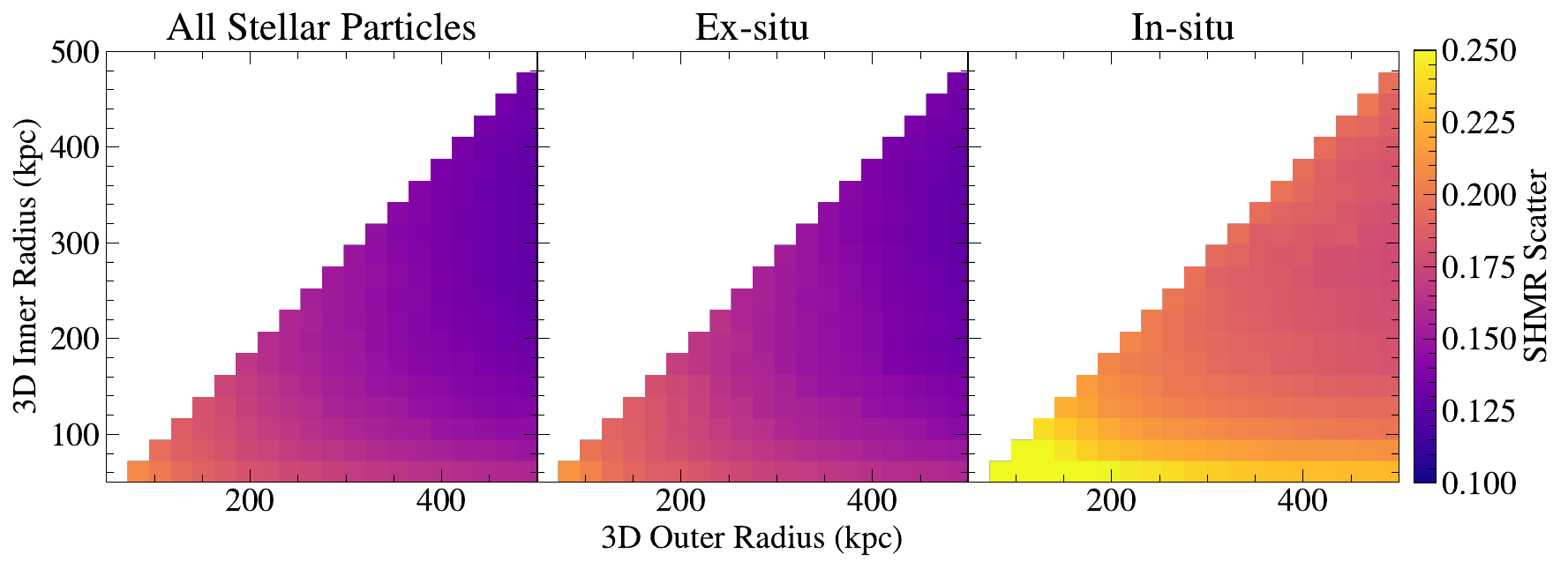}
\caption{Scatter of halo mass at given stellar mass from the different 3D outer stellar mass definitions with different ellipsoidal radii. The inner and outer radii are defined as the ellipsoidal radii of the inner and outer ellipsoids that make up the shell. \textbf{Left:} SHMR scatter using both the in-situ and ex-situ components. \textbf{Middle and right} SHMR scatter using only the in-situ or the ex-situ component. For all three panels, the scatter decreases as we include more particles in the outskirt and exclude more particles around the core. The ex-situ component of the stellar mass is better correlated with the halo mass than the in-situ component. The gradient of SHMR scatter of the in-situ parts shows that not only the outskirt part of the ex-situ stellar mass but also the outskirt part of the in-situ stellar mass is more correlated with halo mass.} 
\label{fig:inner_outer_scatter}
\end{figure*}

The radial trend of increasing ex-situ fraction with radius indeed translates to decreasing the scatter of SHMR. Figure~\ref{fig:inner_outer_scatter} shows the scatter of the SHMR from 3D outer stellar mass definitions with different inner and outer ellipsoidal radii using all stellar particles and the in-situ and ex-situ components separately. The left panel shows that as we go further in the outskirts and exclude more inner parts of the galaxy in our selection, the SHMR scatter gets lower, which agrees with the ex-situ fraction trend. By comparing the middle and right panels, we see again that at a given spatial location, the ex-situ component of the stellar mass is better correlated with the halo mass than the in-situ component. Interestingly, for both ex-situ and in-situ components, the outskirts are better correlated with halo mass than the cores. The gradients in the SHMR scatters for ex-situ only and in-situ only reveal that the high ex-situ fraction is not the only reason why the outskirt stellar mass is a good halo mass proxy. Even for the ex-situ component, the outskirts are better correlated with the halo mass than the cores. These SHMR scatter gradients might be explained by the simple binary labelling of in-situ and ex-situ particles. Not all ex-situ particles are equally correlated with the mass assembly history. Some ex-situ particles accreted at a very early time in the core may have little correlation with subsequent galaxy assembly. This observation gives us more reasons to exclude the galaxy core from our stellar selection.

Having shown that an ellipsoidal shell selection can maximize the ex-situ fraction in 3D and a high ex-situ fraction indeed maps to a low SHMR scatter, we wonder if stars selected with elliptical annuli, which are ellipsoidal shells' 2D natural counterparts, have a similar performance. In the next section, we show that stars selected by 2D elliptical annulus have a comparable or even marginally better performance compared to stars selected by their corresponding 3D ellipsoidal shell. We then explain the good performance of the 2D elliptical annulus selection with a simple and intuitive model.

\begin{figure*}[h]
\includegraphics[width=\textwidth]{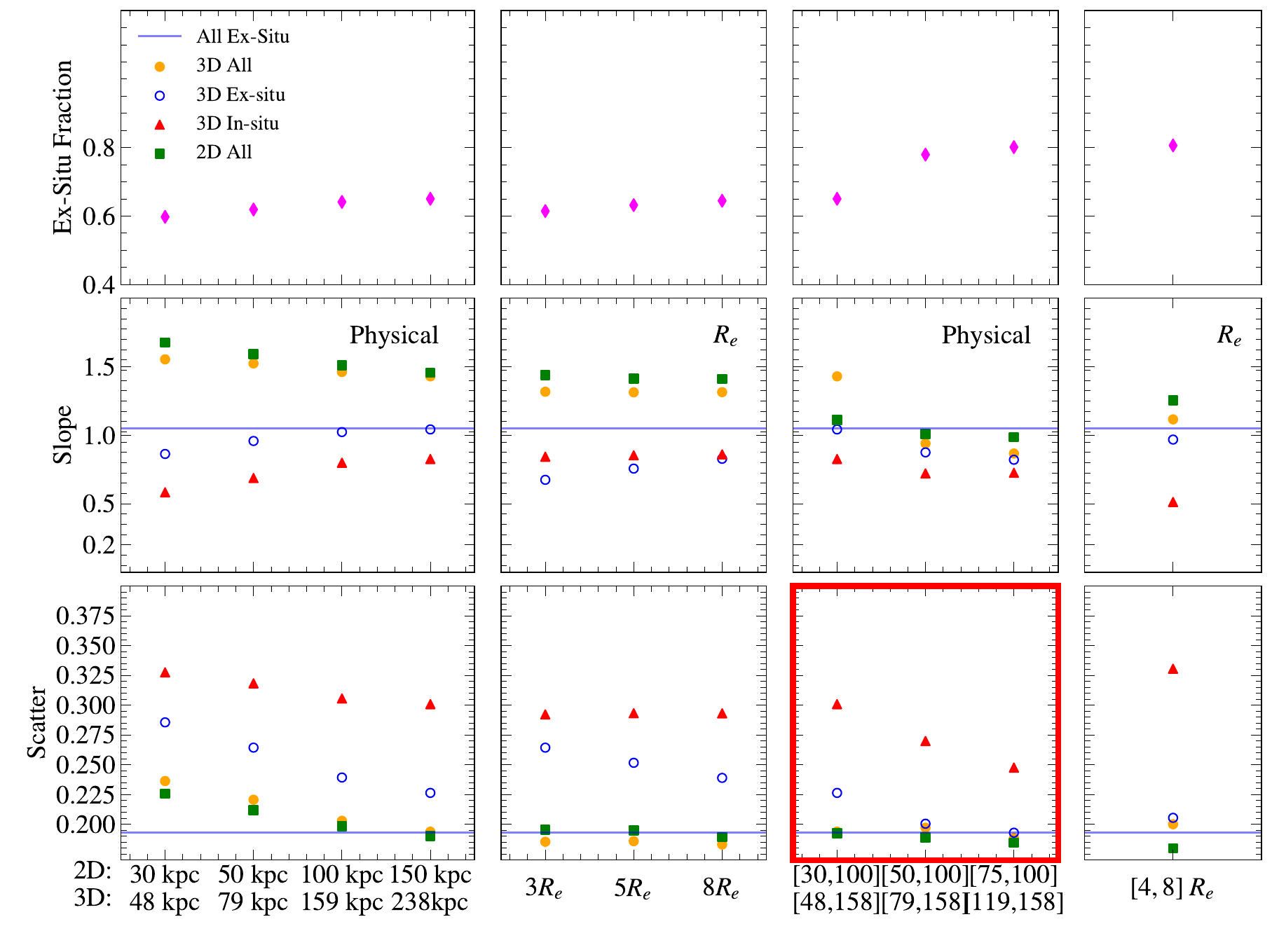}
\caption{\textbf{Top}: ex-situ fraction with different 3D outer stellar mass selections. \textbf{Middle and bottom}: slope and scatter of halo mass at given stellar mass of SHMR with different stellar mass definitions. The size of 2D apertures and annuli is defined as the length of the semi-major axis with elliptical isophotes described in Section~\ref{sec:2d_selection}. The size of 3D apertures and shells is defined following ellipsoidal radii from Eq. \ref{eq:elip_r}. The effective radius $R_e$ in both 2D and 3D is defined as the length of the semi-major axis of the ellipsoid or ellipse that encloses half of the total stellar mass. The 2D physical lengths are mapped to the corresponding 3D physical lengths with Equation~\ref{eq:2to3}. The green squares represent a 2D stellar mass selection. The blue-and-empty circles and the red triangles are the 3D ex-situ only and 3D in-situ only selection. The orange circles represent the 3D stellar mass selection that includes both in-situ and ex-situ components. The blue horizontal solid lines in the second and third rows are the slope and scatter of the SHMR using all ex-situ stellar particles regardless of position. The ex-situ fraction increases as we include more outskirt regions and exclude more inner regions. As the ex-situ fraction increases, the scatter of the SHMR decreases. From the panel highlighted with red borders, we see that the scatters of the physical 2D elliptical annulus selection are comparable to their 3D counterparts for two reasons. First, the 2D elliptical annulus selection also excludes the core of the galaxy and includes more particles along the major axes of the galaxy. Second, the 2D elliptical annulus selection includes more stellar particles in the outskirts since it is a cylindrical selection with no light-of-sight distance limits.}
\label{fig:scatter_slope}
\end{figure*}

\subsection{SHMR scatter, 2D versus 3D}
\label{sec:shmr_scatter}

The top row of Figure~\ref{fig:scatter_slope} shows the mean ex-situ fraction of stellar particles selected with different criteria. The middle and bottom rows show the SHMR scatter and slope with stellar mass from different selections. The size of the 3D apertures and shells are defined in ellipsoidal radii with Equation~\ref{eq:elip_r}. The size of 2D apertures and annuli is defined to be the length of the semi-major axis along the elliptical isophotes described in Section~\ref{sec:2d_selection}. The definition of $R_e$ in both 2D and 3D are the lengths of the semi-major axis of the ellipse or ellipsoid that contain 50\% of the total stellar mass. To compare 2D lengths and 3D lengths, we use Equation~\ref{eq:2to3} to map a 2D length to its 3D counterpart. The blue line in the middle and bottom rows shows the SHMR slope and scatter with all ex-situ stars regardless of the particle position.

The comparison between the ex-situ fraction in the top row and the corresponding SHMR scatter shows that the higher the ex-situ fraction, the lower the scatter of the SHMR. The outskirts of massive galaxies have higher ex-situ fractions, and therefore, the stellar mass in the outskirts are better halo mass proxies compared to the stellar mass of the cores. Furthermore, as we have seen in Figure~\ref{fig:inner_outer_scatter}, both the ex-situ and the in-situ outskirt components are more correlated with halo mass compared to the ex-situ and the in-situ component near the core. Therefore, even if two stellar mass definitions have similar ex-situ fractions, the one that selects particles further from the galaxy center has a lower SHMR scatter. This also explains why the SHMR scatter of the ex-situ outskirt component is sometimes lower than the SHMR scatter of all ex-situ stellar mass, as the former selects a subset of ex-situ stars that contains more information about the assembly history than the average of all ex-situ particles. The highlighted panel in the third row shows that the SHMR scatter of 2D outer stellar mass is comparable with, and sometimes even better, than its corresponding 3D ellipsoidal shell selection. The second and fourth columns show the performance of stellar mass selections based on effective radius $R_e$ for reference. In both 2D and 3D, $R_e$ is defined by the semi-major axis lengths of the ellipse or ellipsoid that contains half of the stellar mass. Although the 2D $R_e$ selection performs slightly better than the 2D physical selection, the difficulty of determining 2D $R_e$ precisely from data may deteriorate its performance. 

However, when it comes to measuring galaxy-galaxy lensing around galaxy clusters, the scatter of the SHMR is not the end of the story. Indeed, we might be concerned that a 2D outer mass selects galaxies with some preferred orientation along the line-of-light. Furthermore, these galaxies could align with the shape of their host dark matter halos. If this were to happen, we might see a ``projection effect" in galaxy-galaxy lensing amplitudes in the sense that the measured signal would not be the same as for galaxies selected with random orientations. For this reason, in the next section, we compare the mean gravitational lensing signal of samples selected by the 2D elliptical annulus stellar mass and the corresponding 3D ellipsoidal shell stellar mass to test if projection introduces unexpected biases.

\begin{figure*} [h]
\includegraphics[width=\textwidth]{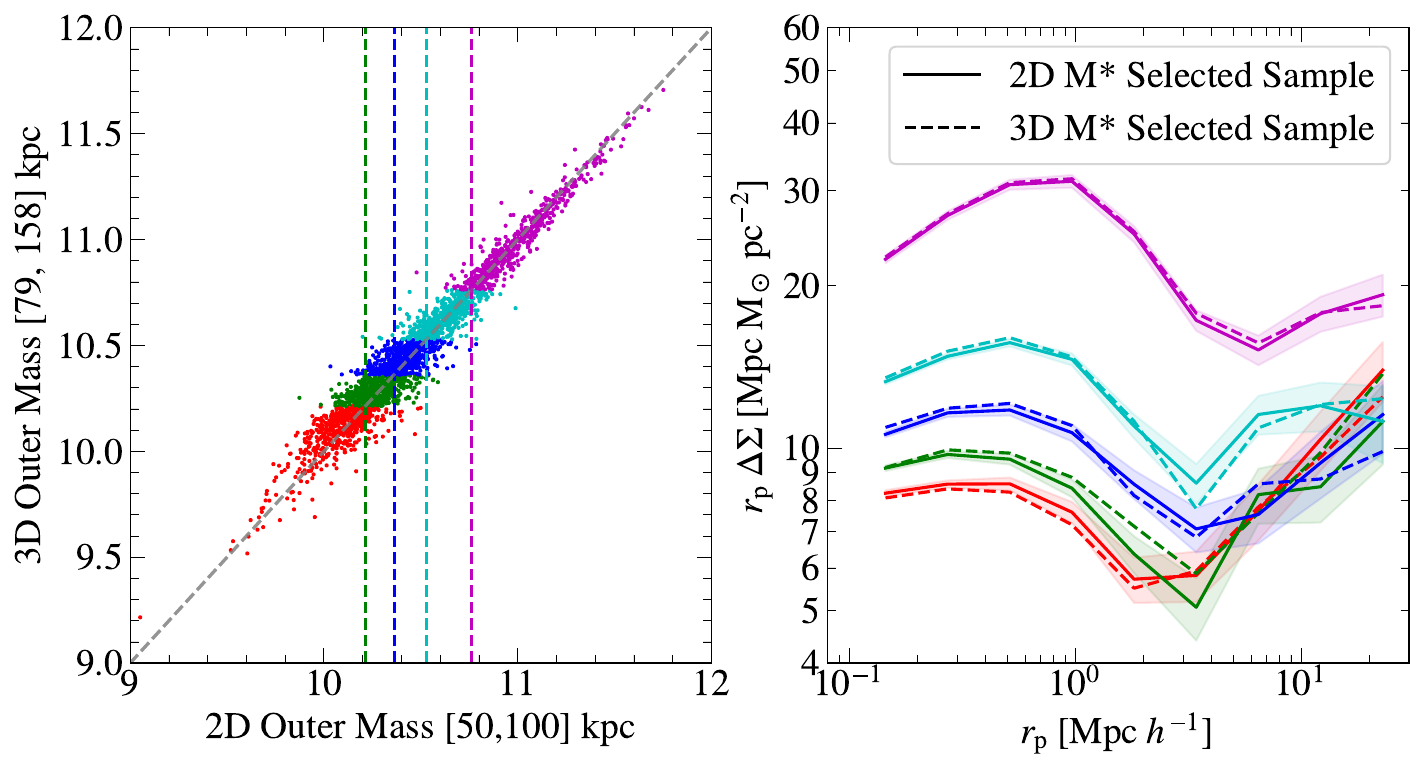}
\caption{\textbf{Left}: Joint distribution of 2D and 3D outer stellar mass. Halos are divided into quartiles based on 3D outer stellar mass, with points color-coded by quartile. For 2D outer stellar mass selection, bins are defined using the same numerical thresholds. The vertical line marks the 2D selection threshold, and the grey dashed line is the 1-to-1 correlation. A strong correlation exists between 2D and 3D outer stellar mass. \textbf{Right}: Mean lensing profiles for 2D and 3D outer stellar mass selected samples, with colors indicating bins from the left plot. Solid (dashed) lines show mean lensing for 3D (2D) selection, with $1\sigma$ bootstrap uncertainty shaded around the solid line. Minimal lensing differences between 2D and 3D selections demonstrate that 2D outer stellar mass selections do not introduce a projection effect in galaxy-galaxy around clusters.}
\label{fig:lensing_comparison}
\end{figure*}

\subsection{Galaxy-galaxy lensing profiles of samples selected by 2D and 3D outer stellar mass}

In this section, we compare the performance of 2D and 3D outer stellar mass as halo mass proxies by comparing the lensing profiles of halos selected by 2D and 3D outer stellar mass. We focus on the $[50,100]$ kpc elliptical annulus stellar mass and its corresponding 3D outer stellar mass selected with $[79, 158]$ kpc ellipsoidal shells. For simplicity, we omit the numerical radius definition in this section and refer to the stellar mass selected by a $[50,100]$ kpc elliptical annulus as ``2D outer stellar mass" and the stellar mass selected by a $[79, 158]$ kpc ellipsoidal shells as ``3D outer stellar mass".
The left panel of Figure \ref{fig:lensing_comparison} shows the joint distribution of the 3D and 2D outer stellar mass. The light gray line marks the 1:1 relation between the 2D and 3D outer stellar mass. The correlation between the 3D outer stellar mass and the 2D outer stellar mass is very high. We further split the sample into four quartiles in 3D outer stellar mass. Each galaxy is colored by its quartile in the 2D vs 3D outer stellar mass plane. We use the same 3D outer stellar mass numerical values to make the equivalent sample selection in the 2D outer stellar mass. The 2D selection cuts are shown as dashed vertical lines with colors corresponding to the quartile of 2D outer stellar mass. The right panel shows the mean $r_{\mathrm{p}} \Delta\Sigma$. The solid lines are the mean lensing of the 3D outer stellar mass selected sample, and the dashed lines are the mean lensing of the corresponding 2D outer stellar mass selected sample. The shaped regions around the solid lines are the 1$\sigma$ uncertainties in the mean lensing profile estimated with $1000$ bootstrap samples . We see that in all stellar mass bins, the mean lensing of the 2D outer stellar mass-selected sample is within 1 $\sigma$ of the mean lensing of its corresponding 3D outer stellar mass-selected sample. Thus, the performance of the 2D outer stellar mass as a proxy for lensing is comparable to that of the 3D outer stellar mass. \emph{This shows that the galaxy-galaxy lensing for 2D outer mass selections is not sensitive to projection effects}. 

In the next section, we attempt to explain the excellent performance of the 2D elliptical annulus selection by analyzing the 3D positions of particles selected in 2D. {\color{red}}

\subsection{3D position distribution of the particles selected in 2D}
In this section, we explain the good performance of 2D outer stellar mass with the 3D particle position distribution selected by a 2D elliptical annulus. The general modeling between a 3D ellipsoid and its 2D projected ellipse requires diagonalizing Schur's complement of the quadratic form that corresponds to the 3D ellipsoid. Given that most galaxies in our samples are highly prolate, we use an approximation in which all galaxies are perfectly prolate with both axis ratios close to zero to get intuitive insight with simple calculations. We name this approximation the line approximation. We discuss the limitations of this approximation in this section and present the full calculation of the 3D spatial distribution stellar particles selected by 2D elliptical annulus in Appendix~\ref{appendix:analytics}. 

\begin{figure} [h]
\centering
\includegraphics[width=0.5\columnwidth]{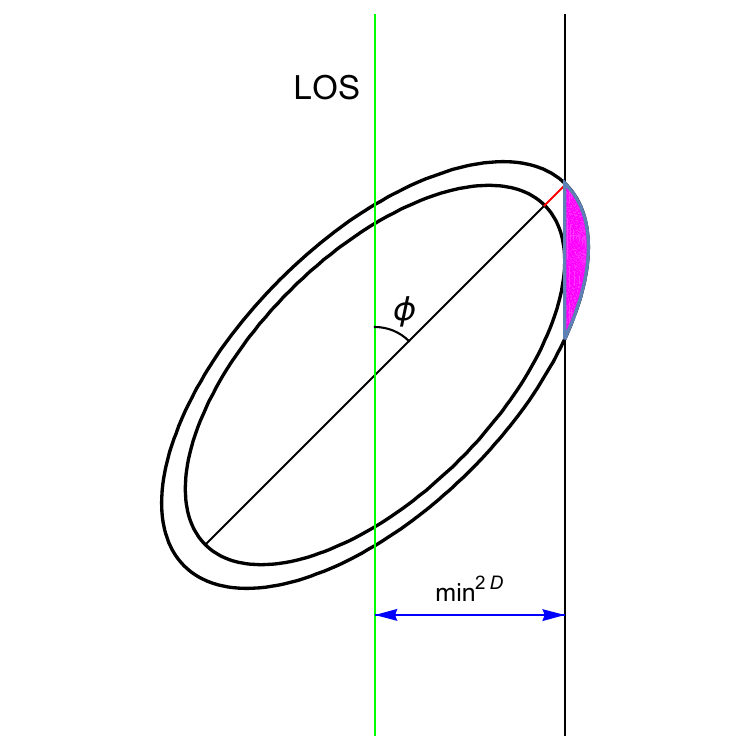}
\caption{Illustration of bias caused by using the line approximation. The ellipse shown is the cross-section of an ellipsoid with the 3D major axis, 2D major axis, and the line of sight vector in the same plane. The blue line is the major axis length of the projected ellipse. In the line approximation, the minimal 3D ellipsoidal distance of particles selected with 2D $[\planemin, \planemax]$ is $\planemin / \sin \phi$, which corresponds to the outer ellipsoid in the figure. However, if the ellipsoid is not perfectly prolate, the ellipsoidal shell that projects to ellipse with semi-major axis $\planemin$ has a semi-major axis length smaller than $\planemin / \sin \phi$ (See Appendix~\ref{appendix:analytics} for a derivation). The minimal ellipsoidal distance of the particle selected is actually slightly smaller than $a \sin \phi$. The true minimal ellipsoidal radius is determined by the inner ellipsoid in the figure. The magenta area in the figure represents particles included in the 2D selection that have an ellipsoidal distance smaller than $\planemin / \sin \phi$. Since most massive central galaxies in our sample are prolate, and this effect is most pronounced when $\phi \approx 45^{\circ}$, it is likely the line approximation is a reasonable way to estimate the 3D ellipsoidal radii of particles selected in 2D.}
\label{fig:tangent_demo}
\end{figure}

In the line approximation, we assume the major axis is much longer than the intermediate axis and the minor axis, and thus $\frac{\solidb}{\solida} = \frac{\solidc}{\solida} \approx 0$. In this approximation, a galaxy with a 3D semi-major axis of $\solida$ can be seen as a line segment with length $2 \times \solida$. If a galaxy's semi-major axis length is $\solida$ and the angle between its major axis and the line of sight direction is $\phi$, for each 3D ellipsoidal shell, its projected shape is an ellipse with semi-major axis length 

\begin{equation}
    \planea = \solida \sin \phi. 
\end{equation}

Then, all the stellar particles selected by $[\planemin,\planemax]$ kpc elliptical annuli have a minimal 3D ellipsoidal radius of at least $\planemin$ kpc since projection always makes $\planea$ smaller than $\solida$. Only when the major axis of the ellipsoid is perpendicular to the line of sight, that is, when $\phi=90^\circ$, the minimal 3D ellipsoidal radius of particles selected is equal to $\planemin$. This is demonstrated in the top row of Figure~\ref{fig:prolate_example} with $\planemin=50$ kpc and $\planemax=100$ kpc. Therefore, in the line approximation, the 2D elliptical annulus selection is very effective in excluding the core of the galaxy.

When galaxies are not perfectly prolate, the line approximations has two limitations. The first one arises when the angle between the major axis of the galaxy and the line of sight is small enough that the length of the major axis of the projected ellipse is no longer set by the projected length of the 3D ellipsoid major axis but the projected length of the 3D ellipsoid intermediate axis. In this case, our conclusion that $\planemin$ is a lower bound of the 3D ellipsoidal distance of the particles selected by $[\planemin,\planemax]$ kpc elliptical annuli remains unchanged. The reason is that if two particles are at the same physical distance from the galaxy center, the one that lies on the intermediate axis always has a greater 3D ellipsoidal distance compared to the particle that lies on the major axis of the galaxy. In the situation where the major axis of the projected ellipse is set by the length of the 3D intermediate axis, the 2D ellipse with semi-major axis length $\planemin$ maps to an ellipsoidal shell with semi-intermediate axis $\solidb$ length $\planemin / \sin \phi$. The semi-major axis length of this ellipsoidal shell $\solida$ is greater than $\planemin / \sin \phi$ since the major axis is greater than or equal to the length of the intermediate axis. Therefore, in this case, the particles selected have greater ellipsoidal distance than $\planemin$, since the ellipsoidal distance of a particle is determined by the length of the major axis of the ellipsoidal isophote the particle lies in,

The second limitation of the line approximation comes from the fact that an ellipsoidal shell of semi-major axis length $a$ and the angle between the major axis and the line of sight $\phi$ does not always project to an ellipse with semi-major axis length $a \sin \phi$. Instead, the projected ellipse sometimes has a semi-major axis that is slightly larger than $a \sin \phi$. Therefore, it's possible to include some particles with an ellipsoidal radius slightly smaller than $\planemin / \sin \phi$ in the 2D elliptical annulus selection. As shown in Figure~\ref{fig:tangent_demo}, the 2D $[\planemin, \planemax]$ selection might include some particles with 3D ellipsoidal radii smaller than $a/ \sin \theta$ in the magenta area. However, as shown in Section.~\ref{sec:morphology}, most of the massive galaxies in our sample are prolate. Moreover, this effect is most pronounced when the angle between the line of sight and the major axis of the ellipsoid is about $45^{\circ}$. Given the shape distribution and the random orientation of our sample, this effect has a minimal impact on our conclusion that the minimal ellipsoidal radius of a sample selected with 2D elliptical annulus is $\planemin$.

\begin{figure*} [h]
\includegraphics[width=\textwidth]{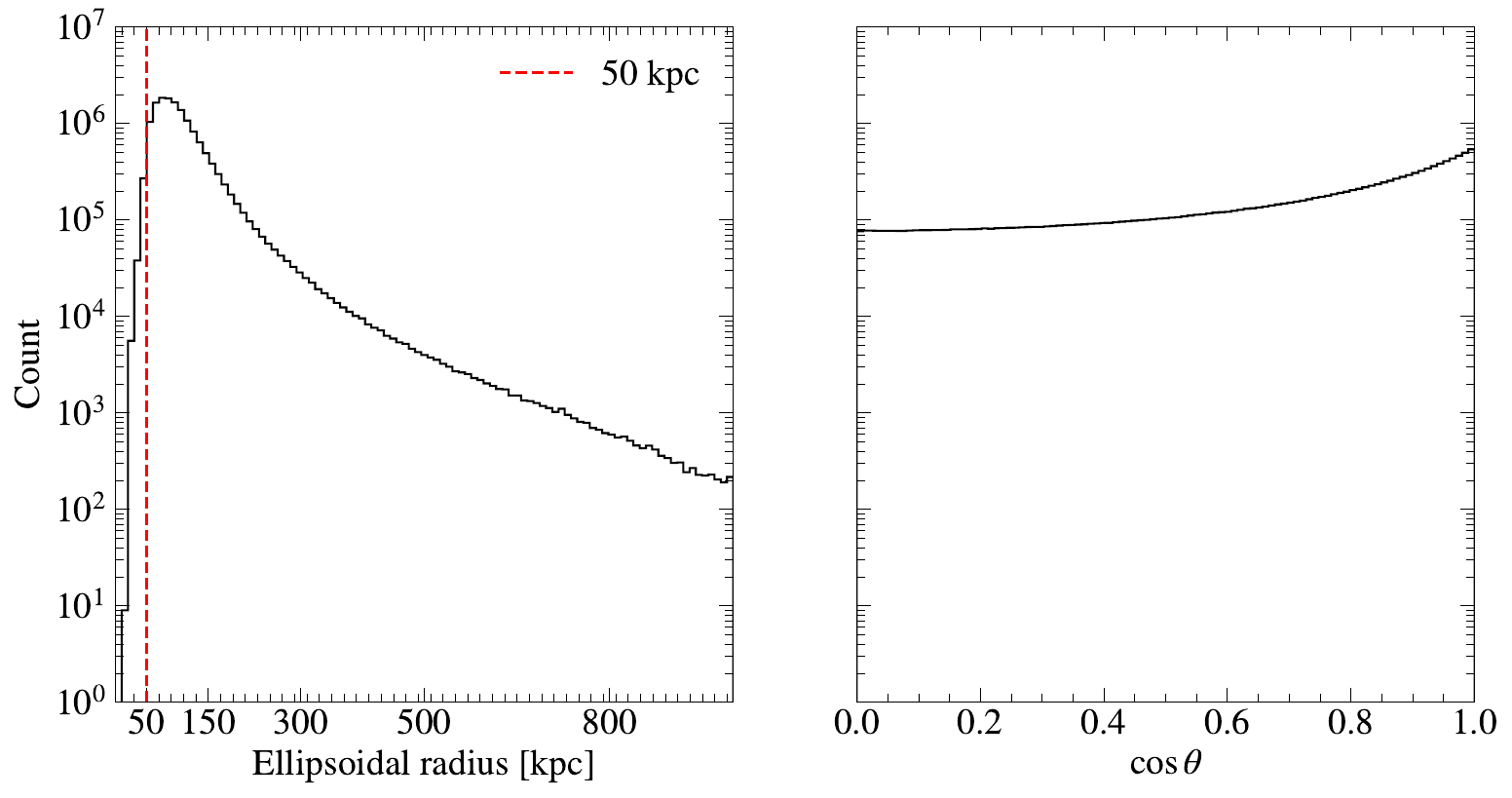}
\caption{Spatial distribution of particles selected in 2D in ellipsoidal polar coordinates. \textbf{Left}: ellipsoidal radii of stellar particles selected within a 2D $[50,100]$~kpc annulus. More than 98\% of the particles have an ellipsoidal radius greater than 50 kpc. The 2D elliptical aperture selection is effective in excluding the galaxy core and in including more particles along the major axis of the galaxy. The small fraction of particles might be from irregularly shaped galaxies for which the 2D measurements and 3D measurements do not have a correspondence. \textbf{Right}: distribution of $\cos \theta$ of the angle $\theta$ between the major axis of the ellipsoid and the stellar particle. The 2D elliptical annulus selection also tends to select particles that are closer to the major axis of the ellipsoid.}
\label{fig:3d_radius_in_50_100}
\end{figure*}

The top row of Figure~\ref{fig:prolate_example} shows the 3D ellipsoidal radii of particles selected by a 2D $[50,100]$ kpc elliptical annuli for the same prolate halo viewed in different directions. We see that, when $\phi \neq 90^\circ$, the minimal ellipsoidal distance of particles is greater than $50$ kpc, and when $\phi = 90^\circ$, the minimal ellipsoidal radii of particles selected is equal to $50$ kpc. The left panel in Figure~\ref{fig:3d_radius_in_50_100} shows the distribution of 3D ellipsoidal radii of all stellar particles selected in 2D $[50,100]$ kpc elliptical annuli. More than 98\% of particles have a 3D ellipsoidal radius greater than 50 kpc. The small fraction of particles with an ellipsoidal radius smaller than $50$ kpc might be caused by irregular galaxy shapes where 3D and 2D measurements have no correspondence. Moreover, in contrast to the 3D ellipsoidal shell case where the ellipsoidal radius of the particles selected has a definitive upper bound, there is a significant fraction of particles with an ellipsoidal radius greater than 100 kpc. This can be explained by the fact that the 2D elliptical annulus selection is a cylindrical selection with no limits in the line of sight distance. This might explain why the 2D outer stellar mass performs slightly better compared to its 3D counterpart since the SHMR scatter decreases with ellipsoidal radius. The right panel of Figure~\ref{fig:3d_radius_in_50_100} shows the distribution of $\cos \theta$ of the angle between the selected particle and the major axes of galaxies. The 2D selection also keeps more particles along the major axes. 

The 2D elliptical annulus selection preserves the advantages of a 3D ellipsoidal shell selection as a halo mass proxy: it is good at excluding the galaxy core and including more particles along the galaxy's major axis. Combined with the ex-situ fraction distribution and the radial SHMR scatter trend we see in Section~\ref{sec:ex_situ_distribution}, the 2D elliptical aperture selection is an optimal 2D selection with minimal SHMR scatter. Moreover, compared with the 3D ellipsoidal shell selection, the 2D elliptical annulus selection has a slight advantage of selecting particles more correlated with the halo mass, since the 2D selection does not have an upper bound on the ellipsoidal radii of the particles selected.

\section{Summary and Conclusions}
\label{sec:summary}

In this paper, we studied the morphology and spatial distribution of ex-situ stars for massive galaxies in IllustrisTNG300. We investigated how stellar mass selections in different 2D and 3D apertures correlate with halo mass. We also asked if 2D stellar mass selections lead to project effects in galaxy-galaxy lensing amplitudes. Our key results are:

\begin{itemize}
\item In IllustrisTNG 300-1, massive central galaxies with stellar mass greater than $10^{11.2} \Msun$ are predominantly non-spherical. 2D light profiles of massive galaxies are, therefore, prone to projection effects.

\item We confirm that, in IllustrisTNG 300-1, the ex-situ fraction of stellar particles in the outskirts of massive galaxies is higher than near the core \citep{rodriguez-gomezStellarMassAssembly2016}. Moreover, there is an anisotropy in the distribution of ex-situ stellar particles: at a given ellipsoidal radius, stellar particles near the major axis of the massive central galaxy are more likely to be of ex-situ origin. 

\item We find that ex-situ stars correlate better with halo mass compared to in-situ stars. Furthermore, we find that both in-situ and ex-situ stars correlate better with halo mass at larger radii. Outskirt mass may contain more stars assembled at late times, thereby correlating better with present day halo mass. 

\item We find that a 3D ellipsoidal shell selection optimally maximizes the ex-situ fraction and minimizes SHMR scatter for two reasons. First, the 3D ellipsoidal shell selection can effectively select stellar particles in the outskirts and exclude stellar particles near the core. Second, the elliptical annuli aperture is more likely to include stellar particles along the major axis of the galaxy. 

\item We find that a 2D outer stellar mass selection with  $r=[50, 100]$ kpc has scatter comparable to the corresponding 3D outer stellar mass, or the stellar mass of all ex-situ stars. Theoretically, it may be possible to use larger 2D elliptical annuli at $r>100$ kpc to get better performance (Leidig et al. in prep.). However, the increasing difficulty of precise background subtraction might prevent us from using larger elliptical annuli. 

\item We find that 2D elliptical annulus selection (the 2D analog to ellipsoidal shell selection in 3D) is the optimal stellar selection for a halo mass proxy in 2D since it preserves the advantages of the ellipsoidal shell selection. Finally, because a 2D selection is projected along the line of sight, it also includes 3D stellar particles in the outer reaches of galaxies beyond a 3D radius of 100 kpc.

\item We find that a 2D outer stellar mass selected in elliptical annuli aperture [50,100] kpc corresponds to its 3D outer stellar mass counterpart selected within [79,158] kpc.

\item Finally, we study potential projection effects in galaxy-galaxy lensing amplitudes. We find that 2D and 3D outer mass selections have similar galaxy-galaxy lensing amplitudes. This suggests that the process of projecting from 3D to 2D does not lead to bias (and does not introduce scatter) for galaxy-galaxy lensing profiles.
\end{itemize}

In this paper, we use only one simulation box, namely IllustrisTNG 300-1. We expect our general conclusion to hold in other hydrodynamical simulations for two reasons. First, we only compare the performance of different stellar selections relative to each other. Second, given that  stellar halos are built up from disrupted satellites we not not expect them to be very sensitive to the specifics of the star formation recipes,  Additionally, we do not consider any observing systematics in the current work to focus on the intrinsic correlation between different stellar mass selections and halo mass. We plan to study mocks with realistic observing conditions to test if the same conclusions still hold. In our current stellar maps, we do not include any galaxies other than the massive galaxy being studied. However, in real data, it is possible a fraction of massive galaxies may be affected by blends. The possibility of blending is likely to increase with the halo mass, which might create a bias in the SHMR. Therefore careful treatment of satellites blended in the stellar halo is required.

The results from the present work underscore the potential of outer stellar mass to select galaxy clusters and to constrain cosmology. In particular, our results show that a 2D outer mass selection is an excellent halo mass proxy, and furthermore, does not introduce projection effects in galaxy-galaxy lensing amplitudes. With the accurate low-surface photometry on data from the Vera C. Rubin Observatory \citep{broughVeraRubinObservatory2020} and Nancy Roman Grace Space Telescope \citep{montesOptimizingRomansHigh2023}, the stellar halos around massive galaxies can be an exciting new venue for precision cosmology.

\acknowledgments

We acknowledge the use of the lux supercomputer at UC Santa Cruz, funded by NSF MRI grant AST 1828315. Parts of the analyses were performed on the YORP cluster administered by the Center for Theory and Computation, part of the Department of Astronomy at the University of Maryland. This material is based upon work supported by the National Science Foundation under Grant No. 2206695 and 2206696. ZCH thanks Aris Alexandradinata and Qiaofeng Liu for helpful discussions on linear transformations.

\appendix

\section{Moment of Inertia Tensor and Ellipsoid}
\label{appendix:moment}

Assume we have a uniform ellipsoid with lengths of semi-axes $a,b,c$ with $a>b>c$. In the coordinate system of the principal axes, the region inside the ellipsoid can be parametrized with 
\begin{equation}
    \frac{x^2}{a^2} + \frac{y^2}{b^2} + \frac{z^2}{c^2} \leq 1
\end{equation}

The volume of the ellipsoid is $V=\frac{4}{3}\pi abc$. The minor-to-major and intermediate-to-major axis ratios are 

\begin{equation}
    q = \frac{c}{a}, s = \frac{b}{a}. 
\end{equation}

We define the moment of inertia tensor to be 

\begin{equation}
    \mathcal{I}_{ij} = \int r_i r_j \; dV.
\end{equation}
We can calculate the components of the moment of inertia tensor to be 
\begin{equation}
    \mathcal{I}=\frac{V}{5}\left(
\begin{array}{ccc}
 a^2 & 0 & 0 \\
 0 & b^2 & 0 \\
 0 & 0 & c^2 \\
\end{array}
\right).
\end{equation}

We define the reduced moment of inertia tensor to be
\begin{equation}
    \mathcal{M}_{ij} = \int \frac{r_i r_j}{r_p^2} \; dV,
\end{equation}
where

\begin{equation}
    r_p=\left(x^2+\frac{y^2}{s^2}+\frac{z^2}{q^2}\right)^{1 / 2}.
\end{equation}
The components of the reduced moment of inertia tensor are
\begin{equation}
        \mathcal{M}=\frac{V}{3a^2}\left(
\begin{array}{ccc}
 a^2 & 0 & 0 \\
 0 & b^2 & 0 \\
 0 & 0 & c^2 \\
\end{array}
\right).
\end{equation}

For a discrete set of $N$ particles in the ellipsoid, we have 
\begin{equation}
    \mathcal{M}_{ij} = \sum_{p=1}^N \frac{r_i r_j}{r_p^2} \frac{V}{N}.
\end{equation}

We don't know the volume of the ellipsoid a priori, so we can only estimate
\begin{equation}
        \frac{\mathcal{M}}{V}=\frac{1}{3a^2}\left(
\begin{array}{ccc}
 a^2 & 0 & 0 \\
 0 & b^2 & 0 \\
 0 & 0 & c^2 \\
\end{array}
\right).
\end{equation}

Therefore, from diagonalizing the reduced moment of inertia tensor, we can get the axis ratio $q,s$ and the direction of principal axes.

\section{Analytical connection between 2D elliptical distance and ellipsoidal distance}
\label{appendix:analytics}

In this Appendix, we characterize the relation between the 2D selected outer stellar mass and its 3D components.

 In general, there is a unique mapping between a 3D galaxy shape and the 2D projected shape at a given projection angle. We assume a 3D galaxy can be described by isodensity ellipsoidal shells with $\rho(\solidr)$ being the radial density function and $\solidr$ is the ellipsoidal radius. The projected density function is $\pi(\planer \mid \solidr,q,s,\theta)$. Each ellipsoidal shell in 3D, in its orthogonal frame, can be described with 

\begin{equation}
    \mathbf{x}^{\prime} \mathbf{\Lambda} \mathbf{x}=1,
\end{equation}

with 
\begin{equation}
    \mathbf{\Lambda}(r)=\left[\begin{array}{lll}
    \frac{1}{\solidr^2} & 0 & 0 \\
    0 & \frac{\solidr^2}{q^2} & 0 \\
    0 & 0 & \frac{\solidr^2}{s^2}
    \end{array}\right].
\end{equation}

The ellipsoid shell's orientation in the observing frame is defined by its axis vectors $\{\mathbf{v_1},\mathbf{v_2},\mathbf{v_3}\}$. We define the line of slight direction to be $\mathbf{v_3}$. The rotation matrix from the ellipsoid's orthogonal frame to the observing frame 

\begin{equation}
    \mathbf{R}^T =\left[\begin{array}{lll}
    v_{11} & v_{12} & v_{13} \\
    v_{21} & v_{22} & v_{23} \\
    v_{31} & v_{32} & v_{33}
    \end{array}\right].
\end{equation}

The ellipsoid shell in the observing frame is $\mathbf{x}^T \mathbf E \mathbf x=1$ with  

\begin{equation}
\mathbf E = \mathbf{R}^T \mathbf{\Lambda} \mathbf{R}.
\end{equation}

The projection of this ellipsoidal shell along $v_3$, $\mathbf F$, can be obtained by computing Schur's complement \citep{schurUberPotenzreihenIm1917},

\begin{equation}
\mathbf F = \mathbf A - \mathbf B \mathbf D^{-1} \mathbf C,
\end{equation}

with 
\begin{equation}
\mathbf{E} = 
\begin{pmatrix}
\mathbf{A} & \mathbf{B} \\[6pt]
\mathbf{C} & \mathbf{D}
\end{pmatrix}
=
\begin{pmatrix}
e_{11} & e_{12} & e_{13} \\[6pt]
e_{12} & e_{22} & e_{23} \\[6pt]
e_{13} & e_{32} & e_{33}
\end{pmatrix},
\end{equation}
where 
\begin{equation}
\mathbf{A} = \begin{pmatrix}
e_{11} & e_{12} \\[6pt]
e_{12} & e_{22}
\end{pmatrix}, \quad
\mathbf{B} = \begin{pmatrix}
e_{13} \\[6pt]
e_{23}
\end{pmatrix}, \quad
\mathbf{C} = \begin{pmatrix}
e_{13} & e_{23}
\end{pmatrix}, \quad
\mathbf{D} = (e_{33}).
\end{equation}

We can diagonalize $\mathbf F$ to get the projected ellipse's axes length and position angle for the projected ellipse. For a diagonalizable $2 \times 2$ matrix, there exist explicit formulae for eigenvalues and eigenvectors. 

Assuming 2D major axis vector is $\mathbf y$, the density at $\planer$ is then

\begin{equation}
    \pi(\planer|a,b,c,\theta) = 2\int_{\solidrtan}^\infty \rho(\solidrtan) \; \mathrm{d} \solidr
\end{equation}

in which $\solidrtan$ is the semi-major axis length of the ellipsoid that projects to the ellipse with semi-major axis $\planer$. $\solidrtan$ can be found using the explicit formulae of 2D eigenvectors and eigenvalues given $q,s,\theta$ and $\planer$.

\bibliographystyle{JHEP}
\bibliography{betterbibtex}

\end{document}